\documentclass[11pt,dvips]{article}
\usepackage{atlas_title,ifthen}

\usepackage{epsfig,wrapfig}

\usepackage{amsmath,amssymb}
\usepackage{times}
\usepackage[varg]{txfonts}
\DeclareMathAlphabet{\mathbold}{OML}{txr}{b}{it}

\usepackage[mathlines,displaymath]{lineno}

\usepackage[numbers,square,comma,sort&compress]{natbib}
\usepackage{hypernat}

\usepackage{caption2}

\usepackage[dvipsnames]{color}
\definecolor{rltred}{rgb}{0.5,0,0}
\definecolor{rltgreen}{rgb}{0,0.3,0}
\definecolor{rltblue}{rgb}{0,0,0.3}

\usepackage[hyperindex,bookmarks,bookmarksnumbered,breaklinks,unicode]{hyperref}
\hypersetup{%
  pdftitle        = {Combining Triggers in HEP Data Analysis},
  urlcolor        = rltblue,       
  urlbordercolor  = 0 0 0.5,
  filecolor       = rltblue,       
  filebordercolor = 0 0 0.5,
  linkcolor       = rltred,        
  linkbordercolor = 0.75 0 0,
  citecolor       = rltgreen,      
  citebordercolor = 0 0.5 0,
  pagecolor       = rltgreen,      
  pagebordercolor = 0 0.5 0,
  menucolor       = rltgreen,      
  menubordercolor = 0 0.5 0,
  colorlinks    = true,
  pdfauthor     = {V. Lendermann, J. Haller, M. Herbst, K. Krueger, H.-C. Schultz-Coulon, R.Stamen},
  pdfsubject    = { },
  pdfkeywords   = {High-Energy Physics, Particle Physics, Experimental Techniques, Trigger, Statistics}
}

\newboolean{twocolumn}
\setboolean{twocolumn}{false}

\newlength{\dinwidth}
\newlength{\dinmargin}
\usepackage[totalwidth=16.0cm,totalheight=24cm]{geometry}

\newenvironment{example}{{\bfseries Example}}{\ensuremath{\bullet}\ }

\def\Nospacing{\itemsep=0pt plus1pt minus1pt \topsep=0pt \partopsep=0pt \parskip=1pt \parsep=0pt}
\let\Otemize =\itemize
\let\Onumerate =\enumerate
\let\Oescription =\description
\renewenvironment{itemize}{\vspace*{-1.5ex}\Otemize\Nospacing}%
                               {\endlist\vspace*{-1.5ex}}
\renewenvironment{enumerate}{\vspace*{-2ex}\Onumerate\Nospacing}
                               {\endlist\vspace*{-2ex}}
\renewenvironment{description}{\vspace*{-2ex}\Oescription\Nospacing}
                               {\endlist\vspace*{-2ex}}

\newdimen\labelwidthi
\leftmargini\labelwidthi

\begin{document}

\makeatletter \def\NAT@space{} \makeatother

\title{Combining Triggers in HEP Data Analysis}

\vspace*{0.1cm}

\author{%
\mbox{Victor Lendermann$^{\rm a}$},
\mbox{Johannes Haller$^{\rm b}$},
\mbox{Michael Herbst$^{\rm a}$},
\mbox{Katja Kr\"uger$^{\rm a}$},
\mbox{Hans-Christian Schultz-Coulon$^{\rm a}$},
\mbox{Rainer Stamen$^{\rm a}$}
}

\vspace*{0.1ex}

\centerline{\begin{minipage}{0.9\textwidth}%
\noindent
$^{\rm a}$\,{\footnotesize Kirchhoff-Institut f\"ur Physik,
Universit\"at Heidelberg, Im Neuenheimer Feld 227, 69120 Heidelberg, Germany}\\[-0.4ex]
$^{\rm b}$\,{\footnotesize Institut f\"ur Experimentalphysik,
Universit\"at Hamburg, Luruper Chaussee 149, 22761 Hamburg, Germany}\\
\end{minipage}}

\begin{abstract}\vspace*{0.1cm}

\noindent
Modern high-energy physics experiments collect data using dedicated complex
multi-level trigger systems which perform an online selection of
potentially interesting events.
In general, this selection suffers from inefficiencies.
A further loss of statistics occurs
when the rate of accepted events is artificially scaled down
in order to meet bandwidth constraints.
An offline analysis of the recorded data must correct for the resulting losses
in order to determine the original statistics of the analysed data sample.
This is particularly challenging when data samples recorded by
several triggers are combined.
In this paper we present methods for the calculation of 
the offline corrections 
and study their statistical performance.
Implications on building and operating trigger systems are discussed.

\end{abstract}

\parindent 0pt

\section{Introduction}\label{s:intro}

Modern high energy collider experiments operating at high interaction rates
rely on complex multi-level trigger systems (see
{\itshape e.g.}~\cite{H1, ZEUS,CDF, D0, ATLAS:2008zzm, CMS:2008zzk})
which select potentially interesting scattering events from large backgrounds.
The selection procedures reduce the initial interaction rates,
 often by several orders of magnitude,
to output rates acceptable for permanent storage.
The recorded events are used in subsequent physics analyses.
The lower level trigger systems are typically built in custom hardware
using information from different detector components.
The higher trigger levels often consist of computer farms performing 
partial or complete event reconstruction which allows the application of
sophisticated decision algorithms.

At each trigger level, events fulfilling the criteria of one or more
independent trigger selections are chosen.
Event losses occur due to inefficiencies of the trigger selections
with respect to the offline analysis.
These inefficiencies result from the coarse
event reconstruction performed within the limited time available at each level.
In addition, the bandwidth restrictions at the different levels may prevent
the recording of all events accepted by certain selections designed to cover
phase space regions with high rates.
The solution applied by the experiments is an artificial 
downscaling of the corresponding event rates.

In an offline data analysis,
the effects of limited efficiency and rate downscaling must be corrected for,
in order to determine the original statistics of the analysed data sample.
This is particularly challenging for analyses of combined event samples
recorded by several independent trigger selections.
Such a combination may be neccessary if the individual trigger selections
cover different regions of the analysed phase space.
Typical cases are:
\ifthenelse{\boolean{twocolumn}}{\vspace*{-1ex}}{}
\begin{itemize}
  \setlength{\itemsep}{0pt plus1pt minus1pt}
  \setlength{\parskip}{1pt}
  \setlength{\parsep}{0pt}
\item Trigger selections based on information from different
detector components,
{\itshape e.g.}\ %
a data analysis relying on trigger
selections using signals from barrel and endcap muon chambers;
\item Trigger selections designed for different kinematic regions,
{\itshape e.g.}\ %
an analysis of events
accepted by several trigger selections
requiring the energy in a calorimeter to exceed different thresholds;
\item Trigger selections sensitive to different objects in the final state,
{\itshape e.g.}\ %
a study of complex final states triggered via
electron, muon and/or jet selections.
\ifthenelse{\boolean{twocolumn}}{\vspace*{-2ex}}{}
\end{itemize}

Ideally a particular combination of trigger selections is already foreseen
at the design stage of the trigger configuration before data taking.
If a combination provides full efficiency for a given signal,
only the downscaling must be corrected for in an offline analysis.
However, for many trigger setups full efficiency cannot be achieved.
In particular, this may be true for analyses unforeseen initially,
in which the necessity of the combination becomes apparent only
in retrospect.

In this paper we provide recipes for the calculation of the
aforementioned corrections.
We discuss their applicability and
statistical performance assuming various trigger setups.
The aim is to achieve the smallest statistical uncertainty.

The paper is organised as follows.
In Sect.\,\ref{s:definitions} basic definitions used throughout the paper
are introduced.
Analyses using event samples recorded via a single trigger selection
are discussed in Sect.\,\ref{s:single}.
Section~\ref{s:onelevel} presents several methods to calculate the corrections
for combined event samples collected with a one-level trigger system.
The corrections of trigger inefficiencies are considered separately.
The recipes are then extended to multi-level trigger systems in
Sect.\,\ref{s:highlevel}.
Finally, the implications for the design and operation of trigger systems
are summarized in Sect.\,\ref{s:rules}.

\section{Basic Ingredients and Definitions}\label{s:definitions}

{\bfseries Trigger selections}.
The decision at each trigger level is based on the fulfillment of requirements 
imposed on event properties, such as
a minimum energy in a calorimeter,  a certain number of tracks
in tracking or muon chambers, or a correct timing of the
signals.
In this paper these pieces of trigger logic are called 
\emph{trigger elements}.
Within one level the trigger elements are combined into logical expressions
(using AND, OR, \dots) which we call
\emph{trigger items}\footnote{Some experiments adopt a different
nomenclature, calling trigger items {\itshape e.g.}\ \emph{subtriggers} or
just \emph{triggers}.}.
A trigger item may, of course, simply consist of a single trigger element.
At each level an event is accepted if it fulfills at least one trigger item.
The rate of events collected by a trigger item can be scaled down
by a \emph{downscale factor} $d$, such that on average only every $d$-th
selected event is kept by the system.
The corresponding downscale procedures can be implemented via
simple counters leading to {\itshape deterministic} downscaling,
or via more sophisticated random selection mechanisms
({\itshape non-deterministic} downscaling).
In multi-level systems, individual trigger items from several levels
are further combined into \emph{chains} (see Sect.\,\ref{s:highlevel}).
Events fulfilling all trigger items within a chain are finally accepted
by the trigger system.

{\bfseries Runs}.
Data at collider experiments are usually collected in event samples
of separate \emph{runs}, in which
stable detector performance and steady running conditions are maintained.
The trigger setup, in particular the downscaling factors
are kept constant within one run, but may vary from run to run
as a reaction to changing conditions, {\itshape e.g.}\ instantaneous luminosity
and background rates.

{\bfseries Trigger bits}.
The states of trigger items in the trigger system are encoded in bits.
We denote by the \emph{raw trigger item bit}:
\begin{equation*}
r_{ij} =
\begin{cases}
\ifthenelse{\boolean{twocolumn}}
{1& \text{if event $j$ is accepted by trigger item $i$}\\[-0.5ex]
  & \text{before downscaling,}\\}
{1& \text{if event $j$ is accepted by trigger item $i$ before downscaling,}\\}
 0& \text{otherwise,}
\end{cases}
\end{equation*}
and by the \emph{actual trigger item bit}:
\begin{equation*}
a_{ij} =
\begin{cases}
\ifthenelse{\boolean{twocolumn}}
{1& \text{if event $j$ is accepted by trigger item $i$}\\[-0.5ex]
  & \text{after downscaling,}\\}
{1& \text{if event $j$ is accepted by trigger item $i$ after downscaling,}\\}
 0& \text{otherwise.}
\end{cases}
\end{equation*}
For the following discussion we assume
that these bits for all trigger levels are stored in the record of each event
and are available for offline data analysis.

{\bfseries Efficiency}.
For an unbiased event sample fulfilling a given analysis selection
the number of events accepted by a raw trigger item 
divided by the original number of events denotes the \emph{efficiency}
$\epsilon$ of this trigger item.
By definition the efficiency depends on the offline selection.

Various techniques for the efficiency determination exist,
which are often specific to certain experiments and physics signals.
A detailed review of these techniques is beyond the scope of this paper.
In general they rely on an event sample collected by a
reference trigger item based on information independent from that
used by the studied trigger item.
Accounting for variations of the efficiency in the phase space,
it is usually determined in bins of certain event parameters $\mathbold{q}$:
\begin{equation} \label{eq:eff}
\epsilon(\mathbold{q}) = \frac{\text{number of events selected by both 
 trigger items}}{\text{number of events selected by reference trigger item}} ~,
\end{equation}
where only events fulfilling the offline event selection are used.
The \emph{actual} bit of the reference trigger item must be set
($a_{ij} = 1$) for all events of the reference sample in order
to ensure their selection by this trigger item, thus 
avoiding any potential bias.
In contrast, for the studied trigger item either the raw or the actual bit
can in principle be used. For the latter, downscale factors have to be taken
into account.
The usage of the raw trigger item however increases the available statistics
by the downscale factor $d$ of this trigger item.
This underlines the importance of storing
the raw trigger item information in the offline event record.
The obtained efficiency distribution is usually fitted by a smooth function,
which can in principle vary from run to run.
In practice, it is determined offline for the entire event sample
or for large subsamples with stable running conditions.

The efficiency of an individual trigger element used within a trigger item
is defined analogously.
For a trigger item consisting of several not fully efficient trigger elements,
the total efficiency can be determined applying
similar considerations as given subsequently
in Sect.\,\ref{ss:eff}
for combinations of several trigger items with inefficiency.

{\bfseries Event weights}.
The recipes presented in this paper provide a \emph{weight} $w_j$
for each event $j = 1, \dots, N$ of the analysed sample
which corrects for the above-mentioned event losses,
such that the original statistics of the analysed event sample
is given by the sum of the weights:
\begin{equation}
N_{\rm ori} = \sum_{j=1}^{N} w_j ~.
\end{equation}
This results in the
\emph{visible} cross section\footnote{The determination of
the true cross section involves further corrections for detector efficiency,
acceptance, etc. which are irrelevant for the present discussion.}
$\sigma$ given by
$\sigma = \dfrac{\sum w_j}{\mathcal L}$
where ${\mathcal L}$ is the integrated luminosity of the event sample.
A non-trivial requirement for each method is that the relative statistical
uncertainty of the cross-section determination should improve with luminosity.

\section{Treatment of a Single Trigger Item}\label{s:single}

If an event sample selected by a single trigger item $i$ is used in an
analysis, {\itshape i.e.}\ $a_{ij} = 1$ for each event $j$,
the weight of the event in run $k$ can be calculated with
\begin{equation} \label{eq:w1single}
 w_{j} = \frac{d_{ik}}{\epsilon_{ik}(\mathbold{q}_{j})} ~,
\end{equation}
where $d_{ik}$ is the downscaling factor for trigger item $i$ in run $k$,
and $\epsilon_{ik}(\mathbold{q}_{j})$ is the efficiency of this trigger item
in this run as a function of a set of event parameters $\mathbold{q}_{j}$.

\begin{example}.
A simple example is given by an analysis using a single trigger item
with a constant downscale factor $d$ and an efficiency $\epsilon$ constant
over the whole parameter space of the physics process under investigation.
In this case the weights of the events $j=1, \ldots, N$ passing the offline
selection criteria,
including the trigger requirement $a_{ij}=1$, are given by $w_j=d/\epsilon$
and the respective visible cross section can be calculated as
$\sigma = \left(\sum w_j\right) / L = (N/{\mathcal L}) \times (d/\epsilon)$.
\end{example}

If the downscaling factors vary strongly from run to run, 
events from runs with high downscale factors in the sample obtain large
weights according to Eq.\,(\ref{eq:w1single}). This leads to
a low statistical significance of the result, especially for differential
distributions, where large statistical errors may occur in certain regions of
phase space. A higher significance is reached if an average weight
over all runs in the whole event sample is used.
With $N$ selected events, with the original number of events $N_{\rm ori}$
and the total cross section of the triggered processes $\sigma$,
the event weight is given by
\begin{equation} \label{eq:w2single}
 w_{j} = \frac{N_{\rm ori}}{N} = \frac{N_{\rm ori}/\sigma}{N/\sigma} =
 \frac{\sum_{k=1}^{N_{\rm runs}}
 {\mathcal L}_k}{\sum_{k=1}^{N_{\rm runs}}
 {\mathcal L}_k \frac{\epsilon_{ik}(\mathbold{q}_{j})}{d_{ik}}} ~,
\end{equation}
where $N_{\rm runs}$ and ${\mathcal L}_k$ are the total number of runs and
the luminosity of the run $k$, respectively.
For a given original number of events $N_{\rm ori}$, {\itshape i.e.}
for a given integrated luminosity of the sample, the averaged weight for
a trigger item depends solely on the total number of events collected
via this item, $N$.
Hence, any optimisation of the downscaling factors during data taking
which leads to a larger collected statistics results in smaller weights
and consequently in a smaller statistical uncertainty.

\begin{figure}[tb]
\centerline{\epsfig{file=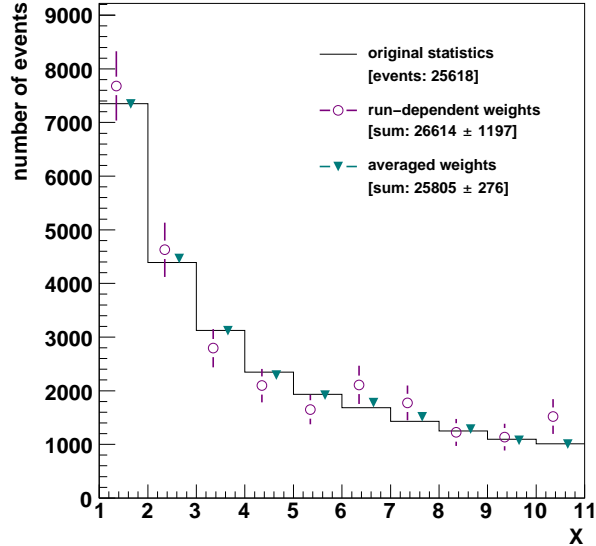,width=0.5\textwidth}}
\caption{Results of a toy Monte Carlo simulation of an analysis relying
on a single trigger item.
The original generated distribution of an example variable $X$ (dashed
histogram) is depicted,
as well as the distribution of triggered events reweighted using
run-dependent (open circles) and averaged weights (closed triangles)
with their respective statistical errors.
 \label{f:oneaver}}
\end{figure}

\begin{example}.
In a toy Monte Carlo (MC) experiment we simulate an analysis relying on
a single trigger item with full efficiency.
The simulated data sample corresponds to 20 runs in which the rate
of the trigger item is scaled down by downscale factors varying from run
to run. Within each run a non-deterministic downscaling procedure is used.
In half of the runs, good running conditions are assumed, such that
the downscale factors are low -- between 1 and 5.
The other 10 runs correspond to bad running conditions affecting the trigger
rate, hence the downscale factors are much larger --
in this example of the order of 100. The run luminosity is varied such that
each run consists of 1000 to 1500 events.
The ratio of the number of events in each run to its integrated luminosity
is smeared using Poissonian statistics.
Figure~\ref{f:oneaver} shows the original distribution of an example variable
$X$, as well as the distributions of triggered events reweighted using the
run-dependent weights of Eq.\,(\ref{eq:w1single})
and the averaged weights of Eq.\,(\ref{eq:w2single}) with their
corresponding uncertainties.
Both methods are able to reproduce the original distribution
but with different statistical performance.
As expected, the application of the averaged weights results in a smaller
statistical uncertainty and thus a much smoother distribution.
This is reflected by the total numbers of events and their uncertainties
obtained with the two methods.~\end{example}

{\bfseries Statistical uncertainty.}
With $N$ selected events,
the statistical uncertainty on the original number of events $N_{\rm ori}$
is given by the standard formula
\begin{equation}
 \delta_{\rm N_{\rm ori}} = \sqrt{\sum_{j=1}^{N} w_j^2}~. \label{eq:sumw2}
\end{equation}
For different sets of $N$ real numbers $w_j$, all having the same sum%
\footnote{The sum of event weights is, of course,
not constant but fluctuates around $N_{\rm ori}$ with the spread given
by Eq.\,(\ref{eq:sumw2}).} $N_{\rm ori}$,
the sum of the squares of these numbers is minimised when all numbers are
equal. This can easily be proven using for instance
the method of Lagrange multipliers or mathematical induction.
Therefore, the application of averaged event weights
(Eq.\,(\ref{eq:w2single})) minimises the statistical uncertainty
 $\delta_{\rm N_{\rm ori}}$.
For the same reason, weight averaging over run ranges improves the result
for all methods of combining triggers described in this paper
({\itshape cf.} Sect.\,\ref{s:onelevel} and \ref{s:highlevel}).

In case of a \emph{deterministic} downscaling procedure,
{\itshape e.g.}\ using
hardware counters, each $d_k$-th event in run $k$ is accepted and the initial
number of events $N_{\rm ori}$ is \emph{exactly} equal to the sum of
event weights and the sum of the counter values $p_k$ at the run ends:
$N_{\rm ori} = \sum w_j+\sum p_k$. Since the second term can be neglected
in the limit of large statistics in individual runs,
one might expect a statistical uncertainty of
$\delta_N = \sqrt{N_{\rm ori}}$. However, this is only true
for the total number of events in the sample accepted by a trigger item.
In the subsequent data analyses, cuts are made and differential
distributions are studied, such that the errors are determined for
subsamples of events. In practice, the sum of event weights in a subsample,
{\itshape e.g.}\ in one bin of a differential distribution, is
\emph{not exactly} equal to the original number of events $N_{\rm ori}$
due to statistical fluctuations of the downscaling procedure
from bin to bin. The sum gives, however, a correct statistical estimate
of $N_{\rm ori}$ within the uncertainty given by Eq.\,(\ref{eq:sumw2}).
For \emph{non-deterministic} downscaling this equation is correct
in all cases.

{\bfseries Systematic uncertainties.}
In a deterministic downscaling procedure, selecting the first or last event
within a downscale interval introduces a systematic error
if the varying value of the downscale counter at the end of each run
is not considered in the analysis.
The relative error for the total number of events
is then of the order of $N_{\rm runs} d / \big(2 \epsilon \sum w_j\big)$,
where $d$ is a typical downscale factor, $\epsilon$ is the average efficiency
and $\sum w_j$ is the sum of weights of all recorded events.
This error is typically negligible except for analyses using
many short runs with large downscale factors.
The uncertainty is further reduced if the selection is performed
in the middle of the downscale interval since, on average, the counter
values at run ends are equally spread around the middle value\footnote{%
Exceptional cases are runs with extremely small statistics selected by
the actual trigger item, {\itshape e.g.}\ resulting from large downscale
factors, low efficiency or short run time, in which
no more than one event per run is selected, and
the downscale counter does not reach on average the middle of the interval.}.
The uncertainty can be completely avoided with a non-deterministic downscaling
procedure, {\itshape e.g.}\ if the downscale system selects events
on a random basis, or if at least a random position of the downscale counter
at each run start is chosen.

\section{Combination of Trigger Items in One-Level Systems}\label{s:onelevel}

In this section we present methods for the calculation of corrections
for event losses in analyses of combined event samples
recorded by several trigger items in a trigger system consisting of only
one level.
The methods are also applicable if the higher trigger levels
accept all events preselected by the first-level trigger items
in the analysed phase space.
The basic concepts discussed here are extended in Sect.\,\ref{s:highlevel}
to the general case of multi-level systems.

\subsection{Division Method}\label{ss:div1}

An obvious approach for a combined analysis on a single trigger level
is the \emph{Division Method}, in which the phase space is
divided into distinct regions in terms of appropriate kinematic variables,
and only events selected by a single actual trigger item are used in each
region, while all other events are not considered. Clearly,
for the smallest statistical uncertainty the trigger item which provides
the largest number of events must be used in the corresponding region.
This division simplifies the task to an analysis of separate samples
each using one trigger item, as described in Sect.\,\ref{s:single}.
The efficiency of the trigger items must be determined individually in the
respective phase space regions, which may introduce a certain complexity
in practice.

\begin{example}.
The phase space is divided into intervals of energy measured in a calorimeter,
in each of which a separate trigger item is used. However,
one of the items includes the requirement of a certain number of tracks
in a tracking chamber. In this case it could be necessary to determine
the efficiency of this trigger item as a function of an appropriate
track-related variable, {\itshape e.g.}\ the number of reconstructed tracks,
for the energy interval in which it is used.~\end{example}

\subsection{Advanced Methods for Fully Efficient Combinations}\label{ss:advanced}

For analyses in which the individual trigger items provide sufficient
statistics in their respective phase space regions, the Division Method
may yield adequate precision. Otherwise, more elaborate approaches
can be used, such as the \emph{Exclusion Method} and the
\emph{Inclusion Method}, described in the following. 

For both methods, a correction for the trigger inefficiency is not
necessary if the chosen
combination of the trigger items is fully efficient in the analysed kinematic
range, as is often the case for combinations designed before data taking.
Note that this does not imply that each individual trigger item is
fully efficient in the whole range, 
but it is sufficient that each event in the original sample
fulfilling the offline selection is
triggered by at least one of the chosen raw trigger items.
The event may then still be rejected by the downscaling procedure.

For this reason we first discuss both methods for the case of full
efficiency. These recipes, though not labelled as in this paper,
have been used in data analyses by the H1 collaboration
({\itshape e.g.}\ in \cite{Adloff:1997da,Adloff:2000qk,Adloff:2002ex})
to correct for downscaling.
Afterwards we present newly developed techniques which include efficiency
corrections.
Finally, we compare the statistical performance of the various methods.

\subsubsection{Exclusion Method for Fully Efficient
Combinations}\label{ss:excl1}

Similarly to the Division Method, the Exclusion Method~\cite{excl}
splits the event sample into subsamples in which single trigger items
are considered. However, the sample is now divided not in terms of
kinematic variables, but according to trigger item bits and downscale factors.
From the set of considered trigger items $i$,
for which the raw trigger has fired ($r_{ij} = 1$) in event $j$ taken
in run $k$, the trigger item $i^*$ with the smallest downscale factor
is chosen:
\begin{equation}  \label{eq:w2excl1}
 i^* : d_{i^*\!k} = \min_{r_{ij}=1}{d_{ik}} ~.
\end{equation}
The weight for the event is then given by
\begin{equation}
w_{jk} = d_{i^*\!k} \, a_{i^*\!j} ~.
\end{equation}
Consequently, the event is rejected, if the actual bit $a_{i^*\!j}$
for the trigger item with the smallest downscale factor is not set.

In case of trigger items with equal downscale factors,
the order, in which the status of the actual bits is checked, is arbitrary,
but must not depend on the status itself.
A simple solution is to define the order once for the whole run range.
A similar prescription holds for every variation of the Exclusion Method 
discussed in the following.

As before, a better statistical significance is reached if weights averaged
over all runs are used ({\itshape cf.} Eq.\,(\ref{eq:w2single})).
In this case, for each considered trigger item $i$, the average weight factor
\begin{equation}  \label{eq:w2excl}
 w'_{i} = \frac{\sum_{k=1}^{N_{\rm runs}}
 {\mathcal L}_k}{\sum_{k=1}^{N_{\rm runs}}
 {\mathcal L}_k \frac{1}{d_{ik}}} ~,
\end{equation}
is calculated once for the whole run range. For all trigger items with the
raw bit $r_{ij} = 1$ in event $j$, the smallest weight factor is then
assigned as the weight to the event, if the corresponding actual bit
$a_{i^*\!j}$ is set, {\itshape i.e.}
\begin{equation}\begin{split}
  \label{eq:w2excl11}
 & i^* : w'_{i^*} = \min_{r_{ij}=1}{w'_{i}} ~,\\
 &w_{j} = w'_{i^*} \, a_{i^*\!j} ~.
\end{split}\end{equation}
Again, the event is rejected, if the corresponding actual bit $a_{i^*\!j}$
is not set\footnote{%
Note that the average weight factor in Eq.\,(\ref{eq:w2excl})
represents an average downscale factor,
and therefore the selection of the minimum in Eq.\,(\ref{eq:w2excl11})
is an analogon of Eq.\,(\ref{eq:w2excl1}).}.

This averaging procedure can only be used if the definitions of all chosen
trigger items remain unchanged during the run range,
as it assumes that if the raw bit is set for an event in a certain run,
it would also be set for an identical event in any other run.
In practice, trigger items may be redefined within the running period,
{\itshape e.g.}\ trigger thresholds may be modified.
For the calculation of event weights the corresponding event sample must
then be split into subsamples with constant definitions.
Consequently, frequent redefinitions of trigger items should be avoided.

\subsubsection{Inclusion Method for Fully Efficient
Combinations}\label{ss:incl1}

In the previously discussed methods the event sample is split into
subsamples, in which the weight calculation for each event is
based on a single trigger item. On the contrary,
in the Inclusion Method~\cite{h1-04-97-517,SchultzCoulon:1999tx}
a \emph{combined} weight based on all considered trigger items is determined
for the \emph{entire} event sample.
For each event, at least one actual trigger item bit from the set of
considered items is required to be set.
Thus, events only triggered by items not considered in the given
analysis are rejected.

The weight calculation is based on the probability to accept the event
after the downscaling procedure.
For a single trigger item $i$ with the downscale factor $d_{ik}$ in run $k$,
this probability for an event $j$ is
\begin{equation}  \label{eq:p1incl}
P_{ijk} = \frac{r_{ij}}{d_{ik}} ~.
\end{equation}
Assuming all downscaling decisions to be independent of each other,
the probability that at least one of the $N_{\rm items}$
trigger items accepts the event is given by
\begin{equation} \label{eq:p2incl}
P_{jk} = 1 - \prod_{i=1}^{N_{\rm items}}
 \left(1 - \frac{r_{ij}}{d_{ik}}\right) ~.
\end{equation}
The run-dependent weight for event $j$ is then
\begin{equation}  \label{eq:w1incl}
w_{jk} = \frac{1}{P_{jk}} ~,
\end{equation}
while the weight averaged over runs is given by
\begin{equation}  \label{eq:w2incl}
 w_{j} = \frac{\sum_{k=1}^{N_{\rm runs}}
 {\mathcal L}_k}{\sum_{k=1}^{N_{\rm runs}} {\mathcal L}_k P_{jk}} ~.
\end{equation}
As for the Exclusion Method, the averaged weight can be used only if the
definition of all chosen trigger items remains unchanged during the run
range, such that it is possible to calculate the triggering probability
of an event in a run different from the one in which it was recorded.

Note, that the assumption of independent downscaling decisions
is not valid in deterministic downscaling systems
containing several (quasi-)identical trigger items\footnote{%
Since identical trigger items accept the same events
their downscaling decisions are made synchronously leading 
to statistical correlation.
Quasi-identical items which select very similar event samples 
follow a synchronous downscaling procedure in parts of the data-taking period.
}.
In this case the above formulae can still be applied if
{\it (i)} the downscaling factors for these items are different and
{\it (ii)} the downscaling factors are coprime integers, or
in general, they are irreducible fractions with coprime enumerators.

\subsubsection{Comparison of the Exclusion and Inclusion
Methods}\label{ss:comp}

While the Division Method and the Exclusion Method only use a fraction
of the total triggered event sample, the Inclusion Method offers the
advantage of using \emph{all} events in the sample and therefore outperforms
the other methods in statistical precision.
For illustration we performed a toy Monte Carlo study comparing the 
Exclusion and the Inclusion Methods.

\begin{figure*}[tb]
\centerline{%
\epsfig{file=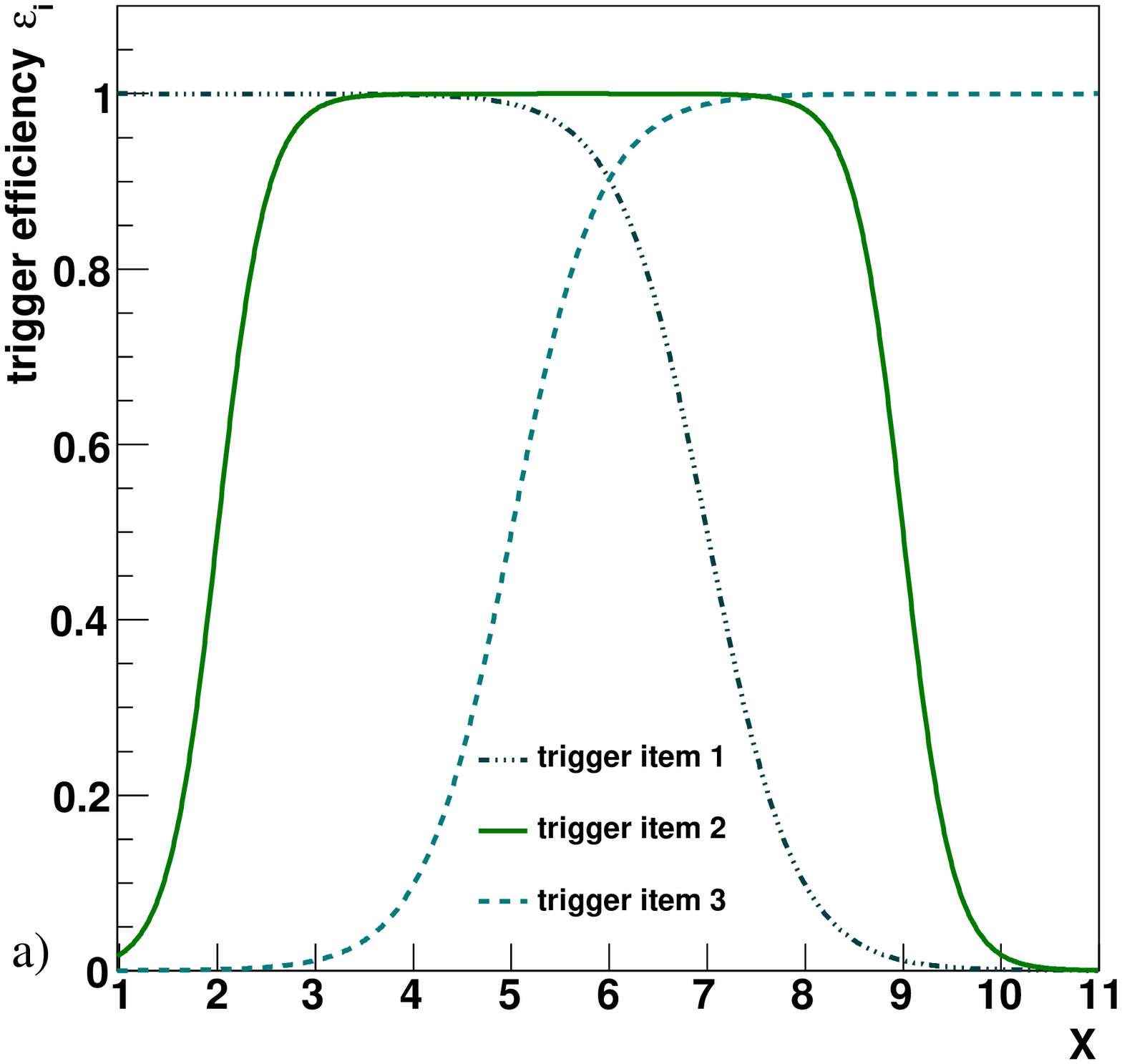,width=0.49\textwidth}%
\epsfig{file=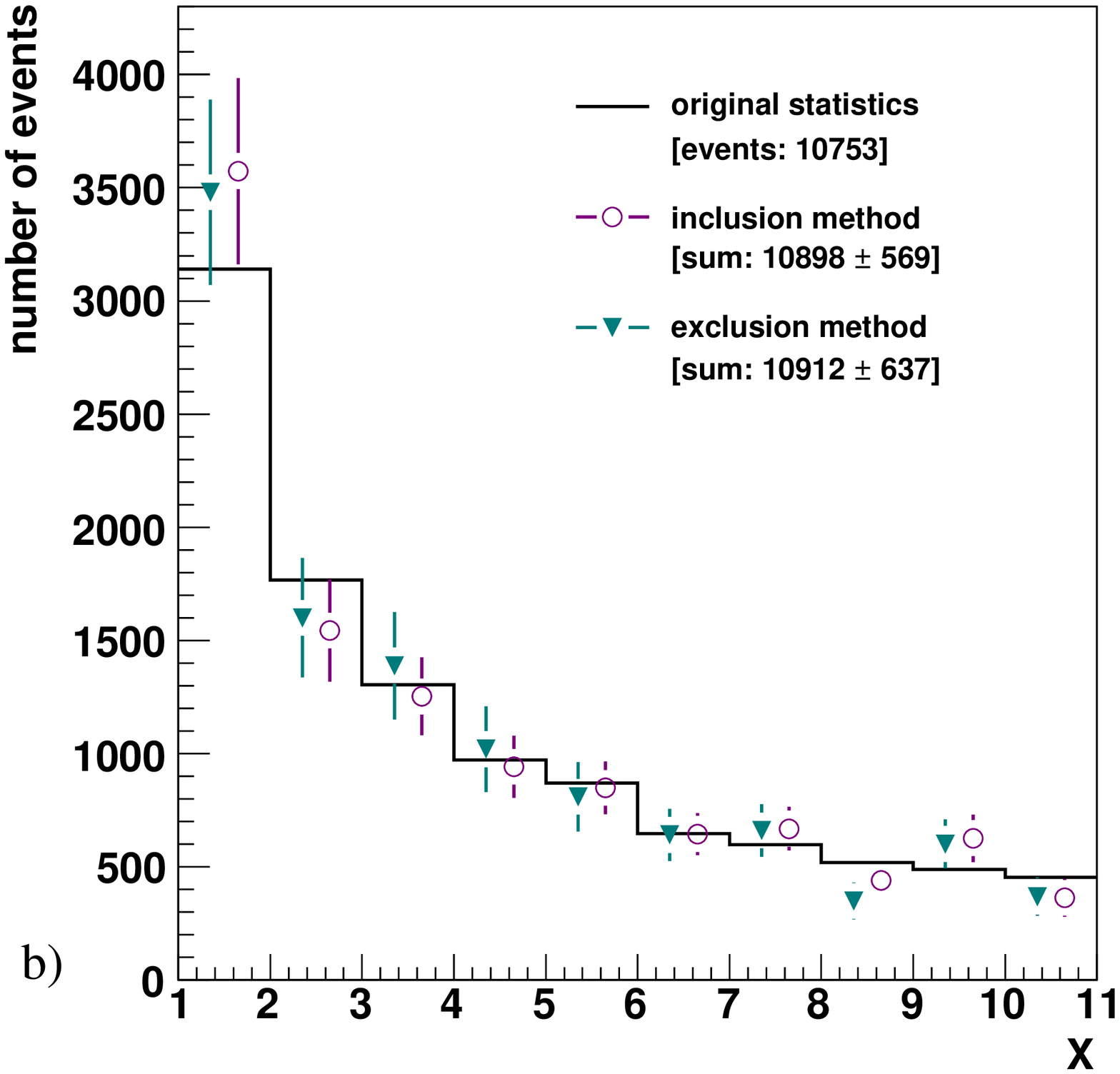,width=0.5\textwidth}%
}%
\caption{a) Assumed efficiencies of the three trigger items
as a function of an event variable $X$ used in the toy Monte Carlo simulation;
b) Original event distribution (dashed line),
as well as the distributions of triggered events reweighted
using the Exclusion Method (closed triangles) and the Inclusion Method
(open circles), both with weights averaged over runs.\label{f:comp}}
\end{figure*}

\begin{example}.
In the MC toy experiment the response of a trigger system with three
items is simulated. The items select events based on the value
of an event variable $X$
 (this could be {\itshape e.g.}\ the energy in a calorimeter).
The assumed efficiencies of the trigger items are shown in
Fig.\,\ref{f:comp}a as a function of $X$.
Each part of the analysed phase space is fully covered by at least one
trigger item, {\itshape i.e.}\ the combination is fully efficient.
An event sample
is simulated corresponding to 20 runs with varying luminosities and
downscale factors. The run luminosity is varied such that each run
consists of 500--600 events. The ratio of the number of events in each
run to its integrated luminosity is spread around a mean value 
following Poissonian statistics. The downscale factors are also varied
from run to run: for the first (second, third) trigger item they are spread
around 50 (40, 20).
In Fig.\,\ref{f:comp}b the original event distribution
is shown as well as the distributions of triggered events reweighted
using the Exclusion and the Inclusion Method with weights averaged over runs.
Both methods provide similar results which reproduce the original
distribution within the statistical uncertainties. As expected,
the Inclusion Method provides a better statistical significance,
as indicated by the error bars and by the error on the total number
of events.~\end{example}

While the Inclusion Method provides by construction a better statistical
precision, the relative improvement with respect to the Exclusion Method 
depends on the concrete experimental set-up and is rather small
in many practical scenarios.
The maximum gain is achieved
if {\itshape (i)} the overlap of efficient regions of the trigger items
is large and {\itshape (ii)} the items have big downscale factors
of similar magnitude
such that the overlap between the event samples actually collected
by the different trigger items is small.

\begin{example}.
Two trigger items with downscale factors $d_1$ and $d_2 \geq d_1$ are
both fully efficient in the analysed phase space,
{\itshape i.e.}\ both raw trigger items fired in all events.
The number of events with the actual trigger item bit 1 or 2 set is given by
$n_1 = N_{\rm ori}/d_1$ and $n_2 = N_{\rm ori}/d_2$, respectively%
\footnote{Statistical fluctuations 
and end-of-run corrections are neglected.},
where $N_{\rm ori}$ is the original number of events.
In total, $N \leq n_1 + n_2$ events are recorded.
With the Exclusion Method, the relative statistical error on $N_{\rm ori}$
is then given by
\begin{equation}
\frac{\delta^{\rm excl}_{N_{\rm ori}}}{N_{\rm ori}} =
 \frac{\sqrt{n_1 d_1^2}}{n_1 d_1} =
 \frac{1}{\sqrt{n_1}} = \frac{\sqrt{d_1}}{\sqrt{N_{\rm ori}}} ~,
\end{equation}
while with the Inclusion Method we get
\ifthenelse{\boolean{twocolumn}}%
{\begin{equation}\begin{split}
&\frac{\delta^{\rm incl}_{N_{\rm ori}}}{N_{\rm ori}} =
\frac{\sqrt{N w^2}}{N w} = \frac{1}{\sqrt{N}}
 = \frac{\sqrt{w}}{\sqrt{N_{\rm ori}}}\\
 &\text{with the weight~~}
w = \frac{1}{\frac{1}{d_1}+\frac{1}{d_2} - \frac{1}{d_1 d_2}} ~.
\end{split}
\end{equation}}%
{\begin{equation}
\frac{\delta^{\rm incl}_{N_{\rm ori}}}{N_{\rm ori}} =
\frac{\sqrt{N w^2}}{N w} = \frac{1}{\sqrt{N}}
 = \frac{\sqrt{w}}{\sqrt{N_{\rm ori}}}
\qquad \text{with the weight} \qquad
w = \frac{1}{\frac{1}{d_1}+\frac{1}{d_2} - \frac{1}{d_1 d_2}} ~.
\end{equation}}
The ratio of the two errors is thus:
\begin{equation}
\frac{\delta^{\rm excl}_{N_{\rm ori}}}{\delta^{\rm incl}_{N_{\rm ori}}}
 = \frac{\sqrt{d_1}}{\sqrt{w}}
 = \sqrt{1 + \frac{d_1}{d_2} - \frac{1}{d_2}} ~.
\end{equation}
The maximum ratio of $\sim$$\sqrt{2}$ is reached if both downscale factors
are large and $d_2 = d_1$ (note $d_2 \geq d_1$ in this example).
For $N_{\rm items}$ trigger items the maximum ratio is $\sqrt{N_{\rm items}}$.%
~\end{example}

\subsection{Additional Corrections for Trigger Inefficiencies}\label{ss:eff}

In the general case of not fully efficient trigger combinations
additional corrections must be performed.
Basically, two conceptually different approaches are possible.
One approach is based on the determination of a single \emph{global} efficiency
for the combination of all involved trigger items in the whole phase space.
This approach has however several drawbacks:
\begin{itemize}
 \setlength{\itemsep}{0pt plus1pt minus1pt}
 \setlength{\parskip}{1pt}
 \setlength{\parsep}{0pt}
\item Since different trigger items depend in general on different event
properties, a global correction will typically be non-universal but specific
for the given data sample with given selection cuts.
Therefore any change of the analysis selection requires
a new determination of the global efficiency correction,
as the mixture of data samples taken by different trigger items may
vary both with cuts and from run to run.
\item 
The efficiency correction is applied on top of the correction for
downscaling, and therefore must be determined for
the combination of \emph{not downscaled} trigger items. If the efficiency
is determined from data, a proper event subsample must be selected
in which the relative contributions of subsamples collected by
different trigger items are the same as for the combination of
the not downscaled items.
\item A determination of the global efficiency from data
may be unfeasible if no trigger item exists which is orthogonal
to all involved trigger items and provides sufficient statistics.
\end{itemize}
For these reasons the determination of a global trigger efficiency
is in many cases only possible using Monte Carlo simulations.
This implies a high level of understanding of the detector
and of the trigger system to be available in such simulations,
which, if at all, is usually reached only after several years of data taking.

An alternative approach for efficiency corrections
is based on a \emph{separate} determination of the efficiency for each
trigger item. This requires modifications of the procedures of weight
calculation, as described in the following.
For the further discussion we assume the efficiency correction function
$\epsilon_{ik}(\mathbold{q})$ to be known for each trigger item $i$
in run $k$.

\subsubsection{Efficiency Correlations}\label{ss:corr}

For the modification of the trigger combination methods with separate
efficiency functions, correlations between trigger efficiencies must be 
considered. Contrary to the downscaling,
trigger efficiencies are not {\itshape a priori} independent,
{\itshape i.e.}\ the efficiency $\epsilon_{i|m}(\mathbold{q})$ of the
trigger item $i$ for events in which a different raw trigger item $m$ has fired
is not necessarily the same as the efficiency $\epsilon_{i}(\mathbold{q})$
for all events.
Correlations can result from technical/instrumental effects or
physical/kinematic event properties.

\begin{example} {\bfseries of technical effects}.
The efficiencies are certainly correlated if the trigger items include the
same inefficient trigger element. They can be correlated if trigger elements
of different trigger items are implemented in the same electronics.
For instance, several trigger items which include elements triggering
on the jet energy differ in the energy thresholds or
in the required number of jets.~\end{example}

\begin{example} {\bfseries of kinematic effects}.
For a trigger item $1$ requiring a certain value of energy in 
a calorimeter and a trigger item $2$ demanding a certain number
of tracks in a tracking chamber, an efficiency correlation 
arises from the physical correlation between the number of tracks 
and the energy.
In such cases the efficiencies can often be defined in an
independent way if they are determined as functions of proper kinematic
variables. 
In this example, the efficiencies determined as a function of 
the calorimeter energy $E$ for the first trigger item and as a function
of the number of tracks $N$ for the second one may be uncorrelated, such that
 $\epsilon_{1|2}(E) = \epsilon_{1}(E)$,
 $\epsilon_{2|1}(N) = \epsilon_{2}(N)$.
The first relation holds if the efficiency of the calorimeter trigger
depends solely on the energy but is independent of the type of particles
depositing the energy. In this case the efficiency in each energy bin is
independent of
the fraction of charged particles in the signal and therefore on the number of
tracks. Similarly, the second relation holds if the efficiency of the
track trigger is  a function of the track multiplicity only and is unaffected
by the track momenta.~\end{example}

\subsubsection{Expected Trigger Item Bit}\label{ss:expected}

\begin{figure}[tb]
\centerline{\epsfig{file=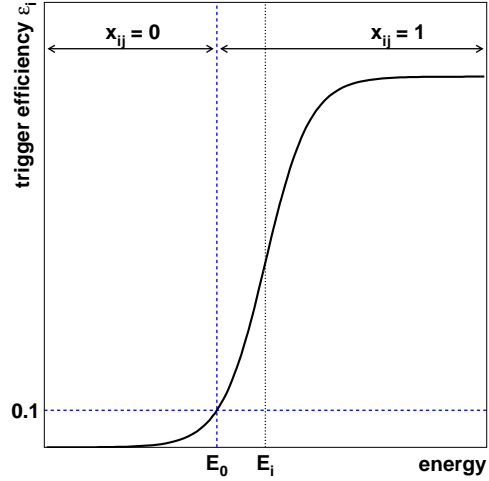,width=0.4\textwidth}}
\caption{Efficiency correction function for an example calorimeter
trigger item with the threshold $E_i$.
Also indicated are the two chosen phase space regions which differ
in the value set for the expected trigger item bit $x_{ij}$.
\label{f:oneeff}}
\end{figure}

In Eq.\,(\ref{eq:eff}) the trigger efficiency is defined with respect to
the offline selection.
For each trigger item we introduce the \emph{expected trigger item bit}
which is set to one if the offline reconstructed event falls into a
specifically chosen region of phase space with significant trigger
efficiency, {\itshape i.e.}\ for which the trigger item 
is expected to fire with sufficiently high probability:
\begin{equation*}
x_{ij} =
\begin{cases}
\ifthenelse{\boolean{twocolumn}}
{1& \text{if event $j$ lies inside the chosen phase space region}\\[-0.5ex]
 & \text{for trigger item $i$,}\\}
{1& \text{if event $j$ lies inside the chosen phase space region
         for trigger item $i$,}\\}
 0& \text{otherwise.}
\end{cases}
\end{equation*}
\begin{example}.
A trigger item $i$ is designed to fire if the energy in a calorimeter
exceeds a certain threshold $E_i$.
Due to the coarse determination of the energy in the trigger,
the efficiency measured as a function of the offline reconstructed energy
is not a step function at $E_i$ but a smoothly
rising Fermi function as shown in Fig.\,\ref{f:oneeff}.
Since the usage of a trigger item in phase space regions where its efficiency
is very small may lead to large event weights (Eq.\,\ref{eq:w1single} or
\ref{eq:w2single}),
one might decide to use this trigger item only at energies $E > E_0$
where its efficiency exceeds a certain value, {\itshape e.g.}\ 10\%.
The expected trigger bit $x_{ij}$ is thus set to one for events with
$E > E_0$ and to zero otherwise.~\end{example}

In practice, a trigger item may consist of a number of trigger elements which
are fully efficient for the analysed signal and of one or a few trigger
elements for which efficiency corrections are determined as functions
of some kinematic variables.
The trigger item is expected to fire if the fully efficient trigger
requirements are fulfilled and the kinematic variables lie in the range for
which the efficiency correction functions are applied in the analysis.

The introduction of the expected trigger bit $x_{ij}$ allows rather
straightforward extensions of the trigger combination methods, where
the raw trigger bit $r_{ij}$ plays nearly the same role with respect to
$x_{ij}$ as the actual trigger bit $a_{ij}$ with respect to $r_{ij}$.
However, while the $r_{ij}$ and $a_{ij}$ bits are set by the trigger system,
the $x_{ij}$ bits are defined in the physics analysis.
As a result, it can happen that the raw and actual trigger bits
$r_{ij}$ and $a_{ij}$ are set, while $x_{ij}$ is not.
Therefore, instead of $r_{ij}$ and $a_{ij}$, one must use
$x_{ij}$ and $x_{ij}a_{ij}$, respectively.
In the above example this means artificially setting
$a_{ij} = 0$ for all events with $E < E_0$.

\subsubsection{Exclusion Method for Combinations of Trigger Items
 with Inefficiencies}\label{ss:excl2}

With the above definitions
the Exclusion Method is easily modified to take efficiencies into account.
The run-dependent weight factor of event $j$ in run $k$ for each 
chosen trigger item $i$, for which the expected bit $x_{ij}$ is set,
is given by
\begin{equation}  \label{eq:w1exeff}
 w'_{ij} = \frac{d_{ik}}{\epsilon_{ik}(\mathbold{q}_{j})} ~.
\end{equation}
Then the trigger item $i^*$ with the smallest weight factor is chosen and
this factor is assigned as the weight to the event if the actual bit
$a_{i^*\!j}$ for the trigger item is set:
\begin{equation}\begin{split}
 &i^* : w'_{i^*\!j} = \min_{x_{ij}=1}{w'_{ij}} ~,\\
 &w_{j} = w'_{i^*\!j} \, a_{i^*\!j} ~.
\end{split}\end{equation}
If the actual bit is not set, the event is rejected.
For weights averaged over runs the expression
\begin{equation}  \label{eq:w2exeff}
 w'_{ij} = \frac{\sum_{k=1}^{N_{\rm runs}}
  {\mathcal L}_k}{\sum_{k=1}^{N_{\rm runs}}
  {\mathcal L}_k \frac{\epsilon_{ik}(\mathbold{q}_{j})}{d_{ik}}}
\end{equation}
is used instead of Eq.\,(\ref{eq:w1exeff}). 
Contrary to the original Exclusion Method
(Eq.\,(\ref{eq:w2excl})), the averaged weights must be calculated
for each event since the efficiency $\epsilon_{ik}$ is in general 
a function of event properties $\mathbold{q}_{j}$.
Furthermore, the modified method allows the usage of
the averaged weights even if the definitions of the chosen 
trigger items change during the run range,
provided the definitions of the expected bits remain unchanged.

In many cases the modified Exclusion Method is a variant
of the Division Method since it divides the phase space into kinematic
regions in each of which one trigger item is used.

\begin{figure}[tb]
\centerline{\epsfig{file=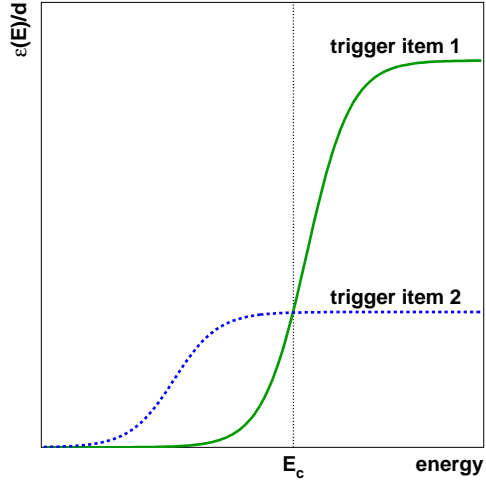,width=0.4\textwidth}}
\caption{Efficiency correction functions divided by the respective
downscale factors for two example trigger items based on the calorimeter
energy.
\label{f:twoeff}}
\end{figure}

\begin{example}.
The analysed data sample is collected by two trigger items
based on the energy $E$ in a calorimeter 
with different thresholds. The trigger item with the higher threshold 
has a smaller downscaling factor. In Fig.\,\ref{f:twoeff} the assumed
efficiency functions for both trigger items divided by the respective 
downscaling factors are shown.
The expected bits for both trigger items are set to one in the whole energy 
range depicted in the figure. The crossing point $E_c$ of the two curves 
divides the phase space, such that for
events with $E >E_c$ ($E < E_c$) only the trigger item with the 
higher (lower) threshold is used. Since the downscale factors and the
efficiencies may vary from run to run, the $E_c$ value may also vary.%
~\end{example}

For this method, possible kinematic correlations of the efficiencies
must be taken into account. In particular, it might be necessary to
redetermine the efficiency functions for the individual phase space regions,
if the efficiencies of the respective trigger items depend on other
variables than those used for the phase space division.

\begin{example}.
Two trigger items, as given in the example of kinematic effects
from Sect.\,\ref{ss:corr}, are used in the analysis.
As a result of the comparison of the ratios $\epsilon_1(E)/d_1$ and
$\epsilon_2(N)/d_2$, the phase space is split into two energy intervals,
such that for energies above (below) a certain value $E_c$, only events
selected by the calorimeter (tracker) trigger item are used.
Due to a possible kinematic correlation between the calorimeter energy
and the number of tracks, the efficiency of the tracker trigger item
may have to be redetermined for the energy range $E < E_c$. Thus
for this trigger item,
one efficiency function $\epsilon_2(N)$ is used to determine the boundary
$E_c$ and another one $\epsilon_{2|E<E_c}(N)$ to calculate the event weight.
The procedure might be improved by iterative redetermination of the
boundary and of the efficiency. Ideally, no redetermination is needed if
the efficiencies for both trigger items are determined as a two-dimensional
function of both $E$ and $N$.
\end{example}

\subsubsection{Inclusion Method for Combinations of Trigger Items
 with Inefficiencies}\label{ss:incl2}

For the Inclusion Method the cases of uncorrelated and 
correlated trigger item efficiencies must be distinguished.
For the former the original procedure can easily be
extended. For each event $j$ in the sample, it is required that from the
chosen list of trigger items, at least one expected trigger item bit $x_{ij}$
and its corresponding actual trigger item bit $a_{ij}$ are set,
{\itshape i.e.}\ %
$x_{ij} a_{ij} = 1$.

The probability that at least one of $N_{\rm items}$ trigger items
accepts the event is given by
\begin{equation} \label{eq:p1ineff}
P_{jk} = 1 - \prod_{i=1}^{N_{\rm items}}
 \left(1 - \frac{x_{ij} \epsilon_{ik}(\mathbold{q}_{j})}{d_{ik}}\right) ~.
\end{equation}
The run-dependent and run-averaged weights are then calculated 
using Eq.\,(\ref{eq:w1incl}) and Eq.(\ref{eq:w2incl}), respectively.

The method for correlated efficiencies is more involved.
For the case of only two trigger items Eq.\,(\ref{eq:p1ineff}) reads
\begin{equation} \label{eq:p2ineff}
P_{jk} = 1 - \prod_{i=1}^{2}
 \left(1 - \frac{x_{ij} \epsilon_{ij}}{d_{ik}}\right) =
\frac{x_{1j} \epsilon_{1j}}{d_{1k}} + \frac{x_{2j} \epsilon_{2j}}{d_{2k}} -
\frac{x_{1j} \epsilon_{1j}}{d_{1k}} \frac{x_{2j} \epsilon_{2j}}{d_{2k}} ~,
\end{equation}
where we use the short-hand notation
$\epsilon_{ij} = \epsilon_{ik}(\mathbold{q}_{j})$.
The first two terms correspond to the respective 
probabilities for each of the two
trigger items to accept the event. The last term gives the overlap probability
that both trigger items accept the event.
This term must be modified to correct for a possible correlation 
of the efficiencies:
\begin{equation} \label{eq:p3ineff}
P_{jk} =
\frac{x_{1j} \epsilon_{1j}}{d_{1k}} + \frac{x_{2j} \epsilon_{2j}}{d_{2k}} -
\frac{x_{1j} \epsilon_{1j}}{d_{1k}} \frac{x_{2j} \epsilon_{2|1j}}{d_{2k}} ~,
\end{equation}
where $\epsilon_{2|1j}$ is the efficiency of trigger item $2$ in event $j$
provided that (raw or actual) trigger item $1$ accepted the event. Note, that
according to Bayes' rule
$\epsilon_{1j} \epsilon_{2|1j} = \epsilon_{2j} \epsilon_{1|2j}$.

\begin{example}.
Two trigger items with downscale factors $d_1$ and $d_2$ have the same
efficiency $\epsilon$ and the expected bits of both items are set
for all events in the analysed data sample. For uncorrelated efficiencies
Eq.\,(\ref{eq:p2ineff}) results in $P_{jk} = \epsilon
 \left(\frac{1}{d_1} + \frac{1}{d_2} - \frac{\epsilon}{d_1 d_2}\right)$.
For fully correlated efficiencies, which would occur if both trigger items
include the same trigger element with efficiency $\epsilon$, the result
of Eq.\,(\ref{eq:p3ineff}) is $P_{jk} = \epsilon
 \left(\frac{1}{d_1} + \frac{1}{d_2} - \frac{1}{d_1 d_2}\right)$,
since in this case $\epsilon_{1|2j} = \epsilon_{2|1j} = 1$ obviously
holds. As expected for the latter case,
the weight calculation factorises
into the correction for downscaling (Eq.\,(\ref{eq:p2incl})) and
the global efficiency correction.~\end{example}

With a dedicated treatment of the overlap probabilities for correlated
efficiencies, the recipe can easily be extended to any number of trigger items.

\subsection{Comparison of Methods with and without Efficiency Corrections}
\label{ss:withwithout}

Though not strictly needed,
the recipes including efficiency corrections can also be used
for trigger item combinations with full efficiency.
This introduces an additional systematic error due to the limited precision
of each efficiency correction, while for the methods without
efficiency corrections, it is sufficient to include only the uncertainty
of the efficiency of one trigger item which is assumed to be fully efficient.
However, if this additional uncertainty is small,
the methods with efficiency corrections may provide a significant gain
of statistical precision.

\begin{example}.
An analysis using the Inclusion Method is based on data samples collected by
two trigger items with the downscale factors $d_1 = 10$ and $d_2 = 1$,
respectively. The first trigger item is fully
efficient; {\itshape i.e.}\ each event in the analysed phase space has
its raw bit set, while the second one has an efficiency $\epsilon = 0.5$.
In practice, such a trigger setup may appear if two trigger items are
based on the same event property with different thresholds.
The trigger item with the lower threshold is more efficient but has a
higher prescale factor. With the original number of events $N_{\rm ori}$,
the number of events $N_a$ which are accepted
only by the actual trigger item 1 and rejected by the raw trigger item 2
is given on average by $N_a = N_{\rm ori} (1 - \epsilon) / d_1$.
The other $N_b$ accepted events have both raw trigger item bits set and
are accepted by at least one of the actual trigger items, such that
$N_b = N_{\rm ori} \epsilon
 \left(\frac{1}{d_1} + \frac{1}{d_2} - \frac{1}{d_1 d_2}\right)$.
In the Inclusion Method without efficiency corrections,
the events of the first and second category obtain the weights $w_a = d_1$
and
$w_b = 1/\left(\frac{1}{d_1} + \frac{1}{d_2} - \frac{1}{d_1 d_2}\right)$,
respectively. The statistical error of $N_{\rm ori}$ is then given by
\ifthenelse{\boolean{twocolumn}}
{\begin{equation}\begin{split}
 & \delta^{\rm incl}_{\rm nocorr} = \sqrt{N_a \, w_a^2 + N_b \, w_b^2}\\
 & ~~~ = \sqrt{N_{\rm ori} (1 - \epsilon) d_1
 + \frac{N_{\rm ori} \, \epsilon}{\frac{1}{d_1}
 + \frac{1}{d_2} - \frac{1}{d_1 d_2}}}
 \approx 2.35 \sqrt{N_{\rm ori}} ~.
\end{split}\end{equation}}
{\begin{equation}
\delta^{\rm incl}_{\rm nocorr} = \sqrt{N_a \, w_a^2 + N_b \, w_b^2} = 
\sqrt{N_{\rm ori} (1 - \epsilon) d_1
 + \frac{N_{\rm ori} \, \epsilon}{\frac{1}{d_1}
 + \frac{1}{d_2} - \frac{1}{d_1 d_2}}}
 \approx 2.35 \sqrt{N_{\rm ori}} ~.
\end{equation}}
On the other hand, if the efficiency corrections are included,
the expected bits can be set to one for all $N_c$ analysed events,
$N_c = N_a + N_b = N_{\rm ori} \left(\frac{1}{d_1} + \frac{\epsilon}{d_2}
 - \frac{\epsilon}{d_1 d_2} \right)$, and thus all
events obtain the same weight $w_c = 1/\left(\frac{1}{d_1} +
 \frac{\epsilon}{d_2} - \frac{\epsilon}{d_1 d_2} \right)$.
The statistical error is then given by
\begin{equation}
\delta^{\rm incl}_{\rm corr} = \sqrt{N_c \, w_c^2} = 
 \sqrt{\frac{N_{\rm ori}}{\frac{1}{d_1} + \frac{\epsilon}{d_2} -
 \frac{\epsilon}{d_1 d_2}}}
 \approx 1.35 \sqrt{N_{\rm ori}} ~.
\end{equation}
The statistical precision is thus improved by a factor of $1.74$.~\end{example}

The reason for the improved performance of the Inclusion Method
is the assignment of equal weights to all events, leading to the
minimisation of the statistical error, as discussed in
Sect.\,\ref{s:single}.

For the Exclusion Method, the introduction of the efficiency corrections
may lead to a gain or loss of statistical precision depending on
the trigger setup.

\begin{example}.
In the above example, the Exclusion Method without
efficiency corrections provides a statistical uncertainty of
\begin{equation}
\delta^{\rm excl}_{\rm nocorr} = \sqrt{N_{\rm ori} (1 - \epsilon) d_1
 + N_{\rm ori} \epsilon d_2} \approx 2.35 \sqrt{N_{\rm ori}} ~,
\end{equation}
while the application of the efficiency corrections gives a smaller uncertainty
\begin{equation}
 \delta^{\rm excl}_{\rm corr} = \sqrt{N_{\rm ori} \frac{d_2}{\epsilon}}
 \approx 1.41 \sqrt{N_{\rm ori}} ~.
\end{equation}
However, for $d_1 = 2$ instead of $10$, the uncertainty
$\delta^{\rm excl}_{\rm corr}$ would be larger than
$\delta^{\rm excl}_{\rm nocorr}$.~\end{example}

The impact on the statistical precision of the Exclusion Method
depends on the interplay of two opposite effects.
On the one hand, the inclusion of the efficiency corrections increases
the weights for individual trigger items and reduces the statistics.
On the other hand, the rejected events may have had even bigger weights
in the calculation without the corrections.

The recipes including efficiency corrections do not require
the knowledge of raw trigger bits and hence might be the only solutions
in case the raw trigger bits are inaccessible in the data analysis.
However this should not be considered as a motivation for skipping
the raw trigger bits in the data aquisition or offline reprocessing steps,
since the efficiency corrections determined from data
can become significantly less accurate (see Sect.\,\ref{s:definitions}).

\section{Combination of Trigger Items in Multi-Level Systems}\label{s:highlevel}

In multi-level trigger systems each trigger item on a particular trigger level
uses as input events accepted by certain 
trigger items of the previous level.
In the most general case, each lower level trigger item provides accepted
events as input to a number of trigger items on the subsequent trigger level,
and each higher level trigger item accepts events from several trigger items
on the lower level. In the following, a sequence of trigger items with
\emph{exactly} one item on each trigger level is referred to as a
\emph{chain}%
\footnote{In the nomenclature of some experiments,
chains are termed \emph{trigger paths}.}.
The general case then corresponds to a collection of many chains,
with potentially large overlap between incorporated trigger items.

All methods described above can be extended to multi-level trigger systems
provided all bits are known at the analysis step
for all chosen trigger items at all trigger levels.
This is not necessarily guaranteed
in modern trigger systems where higher trigger levels run as filter processes
on computer farms. For a better use of the available computing power
and a faster execution on the filter farms, the following mechanisms
are often used:
\begin{itemize}
 \setlength{\itemsep}{0pt plus1pt minus1pt}
 \setlength{\parskip}{1pt}
 \setlength{\parsep}{0pt}
\item \emph{Early-reject mechanism}.
Chains are evaluated in parallel, and the processing of a chain is stopped
as soon as it is clear that the event cannot be accepted by this chain.
In particular, the corresponding algorithms of the chain
on the higher levels are not run
if an actual trigger item bit is not set on a lower trigger level.
\item \emph{Early-accept mechanism}.
At the last trigger level, trigger items are processed sequentially,
and as soon as the decision to accept the event by one item is reached,
the remaining part of the code is not executed.
The downscaling is then either not performed at the last level or
the statements are checked in the order of increasing downscale factors.
\end{itemize}
In such systems the state of the raw and actual bits at the higher levels
remains unknown. Therefore for early-accept systems
the missing trigger information must be calculated in the offline data
processing, where the selection code, the event parameters and conditions
data, such as the alignment and calibration constants used in the online
processing of the event filter, must be available.
For early-reject systems, the information
must be calculated either in the trigger system after a positive trigger
decision or likewise in the offline data processing.

\subsection{Division Method}\label{ss:div3}

The Division Method can easily be extended to multi-level trigger systems.
The analysed phase space is divided into distinct regions in each of which
events are selected by a single trigger chain. The phase space regions
should be chosen such that the highest statistical significance is reached.
Weight factors for each of the levels involved can be calculated
using Eq.\,(\ref{eq:w2single}).
The total event weight is then given by the product of the weight factors
for all trigger levels.

\subsection{Exclusion Method}

\subsubsection{Exclusion Method for Fully Efficient
Combinations}\label{ss:excl3}

In the Exclusion Method for fully efficient configurations the
run-dependent weight factors for each chain $I$
in event $j$ in run $k$ are given by
\begin{equation}  \label{eq:w1excl2}
 w'_{Ij} = \prod_{l=1}^{N_{\rm levels}} d^l_{ik} r^l_{ij} ~~ (i \in I) ~,
\end{equation}
where $N_{\rm levels}$ is the number of trigger levels,
and $r^l_{ij}$ and $d^l_{ik}$ are the raw bits and downscale factors
for the trigger item $i$ on trigger level $l$ belonging to the chain $I$,
respectively. The chain $I^*$ with the smallest non-zero weight factor
is chosen, and this factor is assigned as the weight to the event,
if all actual bits $a^l_{i^*\!j}$ belonging to this chain are set:
\begin{equation}\begin{split}
 &I^* : w'_{I^*\!j} = \min_{w'_{Ij}\neq 0}{w'_{Ij}} ~,
\\
 &w_{j} = w'_{I^*\!j} \prod_{l=1}^{N_{\rm levels}} a_{i^*\!j} ~~ (i^* \in I^*) ~.
 \label{eq:wexcl2}
\end{split}\end{equation}
The event is rejected if one of the actual bits $a^l_{i^*\!j}$ is not set.
For weight factors averaged over runs, Eq.\,(\ref{eq:w1excl2}) is
replaced by
\begin{equation}  \label{eq:w2excl2}
 w'_{Ij} = \frac{\sum_{k=1}^{N_{\rm runs}}
  {\mathcal L}_k}{\sum_{k=1}^{N_{\rm runs}}
  {\mathcal L}_k \prod_{l=1}^{N_{\rm levels}} \frac{1}{d^l_{ik}}}
  \prod_{l=1}^{N_{\rm levels}} r^l_{ij} ~~ (i \in I)  ~.
\end{equation}
While the raw trigger item bits are set separately for each event,
the ratio in front of the product
can be calculated once for the whole run range.

As in the one-level case, frequent redefinitions of trigger items at all
trigger levels should be avoided. In particular, changes of the setups
at different levels should be done simultaneously in order to keep the number
of different run ranges considered in the analysis as small as possible.

\subsubsection{Exclusion Method for Combinations with Inefficiencies}\label{ss:excl4}

For an extension of the Exclusion Method with limited efficiencies,
efficiency correlations between trigger items
not only within one trigger level but also between different levels
must be taken into account.
For example, algorithms on a higher level may not use the full detector
information, but only ``regions of interest'' in the detector 
identified by the lower trigger level.
For such correlations we introduce the \emph{conditional efficiency}
$\epsilon^l_{ik|L}(\mathbold{q}_{j})$ which is the efficiency
of the trigger item $i$ in run $k$ on level $l$ under the condition
that the actual trigger items on certain lower levels
$L$ forming the given chain are set.

The run-dependent weight factor for each chain $I$ is then calculated using
\begin{equation}  \label{eq:w3excl2}
 w'_{Ij} = \prod_{l=1}^{N_{\rm levels}}
 \frac{d^l_{ik}}{\epsilon^l_{ik|(l-1)\dots1}(\mathbold{q}_{j})}
 x^l_{ij} ~~ (i \in I) ~,
\end{equation}
where $\epsilon^l_{ik|(l-1)\dots1}$ indicates the efficiency under
the condition that all corresponding actual trigger items from the lower
levels $(l-1) \dots 1$ fired.
Weight factors averaged over runs are given by
\begin{equation}  \label{eq:w4excl2}
 w'_{Ij} = \frac{\sum_{k=1}^{N_{\rm runs}}
  {\mathcal L}_k}{\sum_{k=1}^{N_{\rm runs}}
  {\mathcal L}_k \prod_{l=1}^{N_{\rm levels}}
  \frac{\epsilon^l_{ik|(l-1)\dots1}(\mathbold{q}_{j})}{d^l_{ik}}}
  \prod_{l=1}^{N_{\rm levels}} x^l_{ij} ~~ (i \in I) ~.
\end{equation}
With the chain weight factors defined according to Eq.\,(\ref{eq:w3excl2}) or
(\ref{eq:w4excl2}), the event weight is then
calculated using Eq.\,(\ref{eq:wexcl2}).

\begin{figure}[tb]
\ifthenelse{\boolean{twocolumn}}%
{\centerline{\epsfig{file=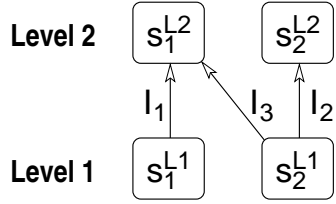,width=0.20\textwidth}}}%
{\centerline{\epsfig{file=twol12big.eps,width=0.27\textwidth}}}%
\caption{Example trigger setup of two levels with two trigger items
on each level, forming three chains.\label{f:twol12}}
\end{figure}

\begin{example}.
In the simplest non-trivial example depicted in Fig.\,\ref{f:twol12},
events in one run are selected by two trigger items
$s^{\rm L1}_1$ and $s^{\rm L1}_2$ on level 1 (L1) and subsequently by two
trigger items $s^{\rm L2}_1$ and  $s^{\rm L2}_2$ on level 2 (L2).
Events accepted by the actual trigger items $s^{\rm L1}_1$ and $s^{\rm L1}_2$
are processed by $s^{\rm L2}_1$, while $s^{\rm L2}_2$ processes only events
accepted by $s^{\rm L1}_2$. Depending on the products of the respective
expected bits  $x^{\rm L1}_1$, $x^{\rm L1}_2$, $x^{\rm L2}_1$, $x^{\rm L2}_2$,
the setup can be considered as three chains:
$I_1 = \{s^{\rm L1}_1 s^{\rm L2}_1\}$, $I_2 = \{s^{\rm L1}_2 s^{\rm L2}_2\}$,
and $I_3 = \{s^{\rm L1}_2 s^{\rm L2}_1\}$.
The weight factors for these chains are
given by the respective downscale factors and conditional probabilities
with obvious notation:
$w_1 = (d^{\rm L1}_1 d^{\rm L2}_1) / 
 (\epsilon^{\rm L1}_1 \epsilon^{\rm L2}_{1|{\text{L1-}}1})$,
$w_2 = (d^{\rm L1}_2 d^{\rm L2}_2) / 
 (\epsilon^{\rm L1}_2 \epsilon^{\rm L2}_{2|{\text{L1-}}2})$,
and
$w_3 = (d^{\rm L1}_2 d^{\rm L2}_1) /
 (\epsilon^{\rm L1}_2 \epsilon^{\rm L2}_{1|{\text{L1-}}2})$.
Events with $x^{\rm L1}_1 x^{\rm L2}_1 = 1$ and with the other products
$x^{\rm L1}_2 x^{\rm L2}_2 = x^{\rm L1}_2 x^{\rm L2}_1 = 0$
get the weight $w_1$. Similarly, events with only
$x^{\rm L1}_2 x^{\rm L2}_2 = 1$ get the weight $w_2$,
and events with only
$x^{\rm L1}_2 x^{\rm L2}_1 = 1$ obtain the weight $w_3$.
Events with only $x^{\rm L1}_1 x^{\rm L2}_2 = 1$ are excluded from the
analysis, since the corresponding chain is not defined.
 For events with $x^{\rm L1}_1 x^{\rm L2}_1 x^{\rm L1}_2 = 1$ and
$ x^{\rm L2}_2 = 0$, the weight factors $w_1$ and $w_3$ are compared.
The smallest weight factor is chosen as the event weight, and only events
with the proper combinatination of actual trigger items
($a^{\rm L1}_1 a^{\rm L2}_1$ for $w_1 < w_3$, or
$a^{\rm L1}_2 a^{\rm L2}_1$ for $w_3 < w_1$) remain in the analysis sample.
In a similar way, events with
$x^{\rm L1}_1 x^{\rm L2}_1 x^{\rm L1}_2 x^{\rm L2}_2 = 1$
are selected or rejected based on the smallest of all three
weight factors.~\end{example}

For the treatment of kinematic correlations,
considerations similar to those discussed in Sect.\,\ref{ss:excl2} apply.
For each chain the efficiencies
$\epsilon^l_{ik|(l-1)\dots1}(\mathbold{q}_{j})$
may have to be redetermined for the corresponding phase space regions.

\subsection{Inclusion Method}

\subsubsection{Inclusion Method for Fully Efficient Combinations}\label{ss:incl3}

The Inclusion Method for fully efficient combinations of chains is described
here following~\cite{h1-04-97-517}
for the case of only two trigger levels.
It can be extended to any number of levels in a straightforward way.

In general, the definition of chains between two trigger levels, L1 and L2,
can be described by the following matrix:
\begin{equation*}
M_{im} =
\begin{cases}
\ifthenelse{\boolean{twocolumn}}
{1 &\text{if L1 trigger item $i$ forms a chain}\\[-0.5ex]
   &\text{with L2 trigger item $m$,}\\}
{1 &\text{if L1 trigger item $i$ forms a chain with L2 trigger item $m$,}\\}
 0 &\text{otherwise.}
\end{cases}
\end{equation*}
Event $j$ is accepted by the trigger system, if at least one of the products
$a_{ij}^{\rm L1} M_{im} a_{mj}^{\rm L2}$ is equal to one.
The probability for the event to be accepted by the downscaling procedure
then depends on the combination of the fired raw trigger items 
$r_{ij}^{\rm L1} M_{im} r_{mj}^{\rm L2}$.
Before discussing the general case of an arbitrary number of items on each
level, we begin with two simple, often occuring and instructive configurations:
\begin{itemize}
 \setlength{\itemsep}{0pt plus1pt minus1pt}
 \setlength{\parskip}{1pt}
 \setlength{\parsep}{0pt}
\item \emph{All-to-1 configuration.} In an analysis based on a single L2
trigger item $m$ 
the probability for an event $j$ in run $k$ to be accepted by L2 trigger item
$m$ is given by
\begin{equation} \label{eq:p1incl2}
P^{\rm L2}_{mjk} = \frac{r^{\rm L2}_{mj}}{d^{\rm L2}_{mk}} ~,
\end{equation}
where $r^{\rm L2}_{mj}$ is the raw bit and
$d^{\rm L2}_{mk}$ the downscaling factor for the L2 trigger item $m$.
The probability for the system to select the event
is given by the product of $P^{\rm L2}_{mjk}$ and the probability
of at least one actual L1 trigger item having fired,
which forms a chain with the L2 trigger item in question
({\itshape cf.} Eq.\,(\ref{eq:p2incl})):
\begin{equation} \label{eq:p2incl2}
P^{\text{L12}}_{mjk} = \left[ 1 - 
  \prod_{i=1}^{N_{\rm L1}} \left(1 -
 \frac{r^{\rm L1}_{ij} M_{im}}{d^{\rm L1}_{ik}}\right) \right]
 \frac{r^{\rm L2}_{mj}}{d^{\rm L2}_{mk}} ~,
\end{equation}
where $r^{\rm L1}_{ij}$ and $d^{\rm L1}_{ik}$ are the raw bit and the downscale
factor for the L1 trigger item $i$, respectively,
and $N_{\rm L1}$ is the number of L1 trigger items.
\item \emph{1-to-all configuration.}
 In an analysis based on a single L1 trigger item $i$
forming chains with several L2 trigger items,
the triggering probability factorises in a similar manner as in
Eq.\,(\ref{eq:p2incl2}):
\begin{equation} \label{eq:p2incl1}
P^{\text{L12}}_{ijk} =  \frac{r^{\rm L1}_{ij}}{d^{\rm L1}_{ik}}
 \left[1 - \prod_{m=1}^{N_{\rm L2}} \left(1 -
 \frac{M_{im} r^{\rm L2}_{mj}}{d^{\rm L2}_{mk}}\right) \right]
 ~,
\end{equation}
with $N_{\rm L2}$ representing the number of L2 trigger items.
\end{itemize}
In the most general case of trigger items entering several chains
on both levels,
the calculation becomes rather involved, since the weight is calculated
based on the raw trigger item bits
independently of the actual trigger item which accepted the event.
However, with the definition of chains (according to the matrix $M_{im}$),
the actual L1 trigger item bits after downscaling influence the decision
to accept the event via an L2 trigger item, and therefore the selection
probabilities of L1 and L2 are correlated and do not factorise. 
The total probability is given by the sum of probabilities
for all combinations (patterns) $S_{\text{\!L1}}$ of actual L1 trigger item bits
that are possible for the raw L1 trigger item setting of the event $j$:
\ifthenelse{\boolean{twocolumn}}
{\begin{equation} \begin{split} \label{eq:p3incl2}
P^{\rm L12}_{jk} = ~ \sum_{S_{\text{\!L1}}} &
\left( \prod_{i \in S_{\text{\!L1}}}
  \frac{r^{\rm L1}_{ij}}{d^{\rm L1}_{ik}} \right)
\left[ \prod_{i \notin S_{\text{\!L1}}}
  \left(1 - \frac{r^{\rm L1}_{ij}}{d^{\rm L1}_{ik}}\right) \right]
  \\
&\cdot
\left\{1 - \prod_{m=1}^{N_{\rm L2}} \left[1 - \left(1 -
  \prod_{i \in S_{\text{\!L1}}} (1 - M_{im}) \right)
  \frac{r^{\rm L2}_{mj}}{d^{\rm L2}_{mk}}
\right] \right\} ~,
\end{split}\end{equation}}
{\begin{equation} \label{eq:p3incl2}
P^{\rm L12}_{jk} = \sum_{S_{\text{\!L1}}}
\left( \prod_{i \in S_{\text{\!L1}}}
  \frac{r^{\rm L1}_{ij}}{d^{\rm L1}_{ik}} \right)
\left[ \prod_{i \notin S_{\text{\!L1}}}
  \left(1 - \frac{r^{\rm L1}_{ij}}{d^{\rm L1}_{ik}}\right) \right] \cdot
\left\{1 - \prod_{m=1}^{N_{\rm L2}} \left[1 - \left(1 -
  \prod_{i \in S_{\text{\!L1}}} (1 - M_{im}) \right)
  \frac{r^{\rm L2}_{mj}}{d^{\rm L2}_{mk}}
\right] \right\} ~.
\end{equation}}
Here, the expression inside the curly braces gives the probability
that an event with a given L1 actual trigger item bit pattern is kept by L2,
while the two products in front give
the probability that this L1 actual bit pattern occurs.
In general, the sum runs over $2^{N_{\rm L1}} - 1$ terms,
which may be a large number.
However, in practice, individual analyses use
only a small number of trigger items
at each level
which makes the usage of Eq.\,(\ref{eq:p3incl2}) feasible.
In addition Eq.\,(\ref{eq:p3incl2}) is simplified for the
following two configurations: 
\begin{itemize}
 \setlength{\itemsep}{0pt plus1pt minus1pt}
 \setlength{\parskip}{1pt}
 \setlength{\parsep}{0pt}
\item \emph{All-to-all configuration.} If several L2 trigger items form
chains with the same set of L1 trigger items
({\itshape i.e.}\ $M_{im} = M_i$ independent on $m$) the probabilities
factorise:
\begin{equation} \label{eq:p4incl2}
P^{\text{L12}}_{jk} = \left[1 - \prod_{i=1}^{N_{\rm L1}}
\left(1 - \frac{r^{\rm L1}_{ij}}{d^{\rm L1}_{ik}}\right) \right]
\left[1 - \prod_{m=1}^{N_{\rm L2}}
\left(1 - \frac{r^{\rm L2}_{mj}}{d^{\rm L2}_{mk}}
\right)\right] ~.
\end{equation}
\item \emph{All-1-to-1-only configuration.} 
For parallel chains, having one separate trigger item on each trigger level,
the matrix $M_{im}$ can be expressed as an identity matrix and 
Eq.\,(\ref{eq:p3incl2}) simplifies to
\begin{equation} \label{eq:p5incl2}
P^{\rm L12}_{jk}
 = 1 - \prod_{i=1}^{N_{\rm items}}
\left(1 - \frac{r^{\rm L1}_{ij}}{d^{\rm L1}_{ik}}
\frac{r^{\rm L2}_{ij}}{d^{\rm L2}_{ik}}\right) ~,
\end{equation}
which is similar to Eq.\,(\ref{eq:p2incl}) for one-level systems.
\end{itemize}
Using the total probability from one of the equations
(\ref{eq:p2incl2})--(\ref{eq:p5incl2}),
the event weight is calculated similarly to the case of one-level systems
({\itshape c.f.} Eq.\,(\ref{eq:w1incl}) or (\ref{eq:w2incl})).
The weight is assigned to the event $j$
if at least one product $a_{ij}^{\rm L1} M_{im} a_{mj}^{\rm L2}$
for the considered trigger items is equal to one.
Otherwise the event is rejected.

For the Inclusion Method with fully efficient trigger configurations
the algorithm of an L2 trigger item must not make use of the 
\emph{actual} L1 trigger item bits, since otherwise the L1 downscaling
enters as an inefficiency of the L2 trigger item and the configuration
is not fully efficient.
In particular, in trigger systems with early-reject mechanism,
one may be tempted to set the higher level raw trigger bit to zero
if the corresponding actual bits at the lower level are unset.
This leads however to wrong weight calculation since
this is equivalent to the inclusion of the lower level actual bits
into the algorithm of the higher level.
On the contrary, the usage of the \emph{raw} L1 trigger item bits 
in L2 algorithms is allowed.

\subsubsection{Inclusion Method for Combinations
 with Inefficiencies}\label{ss:incl4}

Uncorrelated inefficiencies can be included in the same way
as for the one-level system. In Eq.\,(\ref{eq:p2incl2})--(\ref{eq:p5incl2})
the L1 and L2 raw trigger bits must be replaced by the products
of the respective expected bits and efficiencies.
{\itshape E.g.}\ the general expression~(\ref{eq:p3incl2}) is modified to
\begin{equation} \label{eq:p6incl2}
\begin{split}
P^{\rm L12}_{jk} \ifthenelse{\boolean{twocolumn}}{&}{}
= \ifthenelse{\boolean{twocolumn}}{}{~} \sum_{S_{\text{\!L1}}}
\ifthenelse{\boolean{twocolumn}}{}{&}
\left( \prod_{i \in S_{\text{\!L1}}} \frac{x^{\rm L1}_{ij}
  \epsilon^{\rm L1}_{ik}(\mathbold{q}_{j})}{d^{\rm L1}_{ik}}
  \right)
\left[ \prod_{i \notin S_{\text{\!L1}}}
  \left(1 - \frac{x^{\rm L1}_{ij}
  \epsilon^{\rm L1}_{ik}(\mathbold{q}_{j})}{d^{\rm L1}_{ik}}\right) \right]\\
&\cdot
\left\{1 - \prod_{m=1}^{N_{\rm L2}} \left[1 - \left(1 -
  \prod_{i \in S_{\text{\!L1}}} (1 - M_{im}) \right)
  \frac{x^{\rm L2}_{mj}
  \epsilon^{\rm L2}_{mk}(\mathbold{q}_{j})}{d^{\rm L2}_{mk}}
\right] \right\} ~,
\end{split}
\end{equation}
where $x^{\rm L1}_{ij}$, $x^{\rm L2}_{mj}$ are the expected trigger item bits,
and $\epsilon^{\rm L1}_{ik}(\mathbold{q}_{j})$,
$\epsilon^{\rm L2}_{mk}(\mathbold{q}_{j})$ are the efficiency correction
functions for L1 trigger item $i$ and L2 trigger item $m$, respectively.

Efficiencies correlated between trigger items of one level and between
different levels
can be treated in a way similar to Sect.\,\ref{ss:incl2}.
However, the treatment of correlations between different levels must
take into account, whether the conditional efficiencies depend
on the raw or actual trigger items from lower levels.
In case of a dependence on the raw bits, each pattern of actual
trigger items has to be split into the sum of subpatterns with all possible
raw trigger item configurations and conditional efficiencies specific for each
subpattern have to be applied.

\begin{example}.
The example setup of 2$\times$2 trigger items forming three chains
discussed in Sect.\,\ref{ss:excl4} and depicted in Fig.\,\ref{f:twol12}
cannot be reduced to an all-to-1, 1-to-all, all-to-all or all-1-to-1-only
configuration. Hence,
Eq.\,(\ref{eq:p6incl2}) has to be applied giving the probability
\begin{equation} \label{eq:p7incl2}
\begin{split}
P^{\rm L12} = ~ &
\frac{x_1^{\rm L1}\epsilon_1^{\rm L1}}{d_1^{\rm L1}}
 \left(1 - \frac{x_2^{\rm L1}\epsilon_2^{\rm L1}}{d_2^{\rm L1}}\right)
 \frac{x_1^{\rm L2} \epsilon_1^{\rm L2}}{d_1^{\rm L2}}\\
&+
\frac{x_2^{\rm L1}\epsilon_2^{\rm L1}}{d_2^{\rm L1}}
\left(
\frac{x_1^{\rm L2}\epsilon_1^{\rm L2}}{d_1^{\rm L2}} +
\frac{x_2^{\rm L2}\epsilon_2^{\rm L2}}{d_2^{\rm L2}} -
\frac{x_1^{\rm L2}\epsilon_1^{\rm L2}}{d_1^{\rm L2}}
\frac{x_2^{\rm L2}\epsilon_2^{\rm L2}}{d_2^{\rm L2}}
\right) ~.
\end{split}
\end{equation}
The first summand gives the probability that the L1 actual trigger item
$s_1^{\rm L1}$ accepts the event, while the L1 actual trigger item
$s_2^{\rm L1}$ rejects it, and multiplied by the probability 
that the event is then accepted by the L2 actual trigger item $s_1^{\rm L2}$.
If the efficiencies of the items $s_1^{\rm L1}$ and $s_2^{\rm L1}$ on level 1
are correlated, $\epsilon_2^{\rm L1}$ in this summand must be replaced by the
correlated efficiency $\epsilon_{2|1}^{\rm L1}$ for the L1 trigger item 
$s_2^{\rm L1}$ to accept the event, provided the L1 trigger item $s_1^{\rm L1}$
also accepts the event.
If the efficiency of the L2 trigger item $s_1^{\rm L2}$ is conditional and
depends on the raw trigger item bits $r_1^{\rm L1}$ and $r_2^{\rm L1}$, 
then this summand has to be split into two terms
corresponding to the cases that the L1 raw trigger item $s_1^{\rm L2}$
should or should not have fired in the event:
\begin{equation} \label{eq:p8incl2}
\begin{split}
 &\frac{x_1^{\rm L1}\epsilon_1^{\rm L1}}{d_1^{\rm L1}}
 \left(1 - \frac{x_2^{\rm L1}\epsilon_2^{\rm L1}}{d_2^{\rm L1}}\right)
 \frac{x_1^{\rm L2} \epsilon_1^{\rm L2}}{d_1^{\rm L2}} =
\ifthenelse{\boolean{twocolumn}}{\\ &}{}
\frac{x_1^{\rm L1}\epsilon_1^{\rm L1}}{d_1^{\rm L1}}
 \left[x_2^{\rm L1}\epsilon_2^{\rm L1} \left(1 - \frac{1}{d_2^{\rm L1}}\right)
 + (1 - x_2^{\rm L1}\epsilon_2^{\rm L1})\right]
 \frac{x_1^{\rm L2} \epsilon_1^{\rm L2}}{d_1^{\rm L2}}
\ifthenelse{\boolean{twocolumn}}{=\\ &}{\\ &=}
\frac{x_1^{\rm L1}\epsilon_1^{\rm L1}}{d_1^{\rm L1}}
 x_2^{\rm L1}\epsilon_2^{\rm L1} \left(1 - \frac{1}{d_2^{\rm L1}}\right)
\frac{x_1^{\rm L2} \epsilon_1^{\rm L2}}{d_1^{\rm L2}}
 + \frac{x_1^{\rm L1}\epsilon_1^{\rm L1}}{d_1^{\rm L1}}
(1 - x_2^{\rm L1}\epsilon_2^{\rm L1})
 \frac{x_1^{\rm L2} \epsilon_1^{\rm L2}}{d_1^{\rm L2}} ~.%
\ifthenelse{\boolean{twocolumn}}{\hspace*{-0.8cm}}{}
\end{split}
\end{equation}
In the first summand the efficiency $\epsilon_1^{\rm L2}$
has to be replaced by the conditional efficiency
 $\epsilon_{1|{\text{L1-}}12}^{\rm L2}$
of the L2 trigger item $s_1^{\rm L2}$ for the case that both L1 raw
trigger items fired. Similarly in the second summand,
$\epsilon_1^{\rm L2}$ has to be replaced by the conditional efficiency
$\epsilon_{1|{\text{L1-}}1\not{2}}^{\rm L2}$ of the L2 trigger item 
$s_1^{\rm L2}$ for the case that only the L1 raw trigger item $s_1^{\rm L1}$
fired.

The second summand in Eq.\,(\ref{eq:p7incl2}) can be treated similarly.
It gives the probability that the L1 actual trigger item $s_2^{\rm L1}$ and
subsequently at least one of the two L2 actual trigger items accept the event.
If the efficiencies of the L2 trigger items depend on the
raw trigger item bit $r_1^{\rm L1}$, this summand again has to be split
into two terms corresponding to the probabilities that this bit is set or
not set in the event:
\begin{equation} \label{eq:p9incl2}
\begin{split}
 &\frac{x_2^{\rm L1}\epsilon_2^{\rm L1}}{d_2^{\rm L1}}
\left(
\frac{x_1^{\rm L2}\epsilon_1^{\rm L2}}{d_1^{\rm L2}} +
\frac{x_2^{\rm L2}\epsilon_2^{\rm L2}}{d_2^{\rm L2}} -
\frac{x_1^{\rm L2}\epsilon_1^{\rm L2}}{d_1^{\rm L2}}
\frac{x_2^{\rm L2}\epsilon_2^{\rm L2}}{d_2^{\rm L2}}
\right) =\\
 &\frac{x_2^{\rm L1}\epsilon_2^{\rm L1}}{d_2^{\rm L1}}
\left(
\frac{x_1^{\rm L2}\epsilon_1^{\rm L2}}{d_1^{\rm L2}} +
\frac{x_2^{\rm L2}\epsilon_2^{\rm L2}}{d_2^{\rm L2}} -
\frac{x_1^{\rm L2}\epsilon_1^{\rm L2}}{d_1^{\rm L2}}
\frac{x_2^{\rm L2}\epsilon_2^{\rm L2}}{d_2^{\rm L2}}
\right) x_1^{\rm L1} \epsilon_1^{\rm L1} ~ +\\
 &\frac{x_2^{\rm L1}\epsilon_2^{\rm L1}}{d_2^{\rm L1}}
\left(
\frac{x_1^{\rm L2}\epsilon_1^{\rm L2}}{d_1^{\rm L2}} +
\frac{x_2^{\rm L2}\epsilon_2^{\rm L2}}{d_2^{\rm L2}} -
\frac{x_1^{\rm L2}\epsilon_1^{\rm L2}}{d_1^{\rm L2}}
\frac{x_2^{\rm L2}\epsilon_2^{\rm L2}}{d_2^{\rm L2}}
\right) (1 - x_1^{\rm L1} \epsilon_1^{\rm L1}) ~.
\ifthenelse{\boolean{twocolumn}}{\hspace*{-0.5cm}}{}
\end{split}
\end{equation}
In each term of the sum the efficiencies of the L2 trigger items
have to be replaced by the respective conditional ones.
If the L2 trigger item efficiencies are correlated to each other,
the expressions in parentheses have to be modified,
as shown in Eq.\,(\ref{eq:p3ineff}).~\end{example}

In general, if the efficiencies are correlated both within one level and
between different levels,
a significant number of different correction functions may have to be
determined for each trigger item.
One should note that even if some of the used trigger items from different
trigger levels are not combined into a chain, their decisions may be
correlated and hence conditional efficiencies may have to be used.
For instance, the trigger items $s_1^{\rm L1}$ and $s_2^{\rm L2}$
in the above example may be correlated and thus the conditional efficiencies
$\epsilon_{2|{\text{L1-}}12}^{\rm L2}$ and
$\epsilon_{2|{\text{L1-}}\not{1}2}^{\rm L2}$ may differ.

\section{Implications for Design and Operation of Trigger
 Systems}\label{s:rules}

The various methods presented in this paper have consequences
for the design and operation of trigger systems.
Some non-trivial rules are summarised in the following:
\begin{enumerate}
  \setlength{\itemsep}{1pt plus1pt minus1pt}
  \setlength{\parskip}{0pt}
  \setlength{\parsep}{0pt}
\item The raw trigger item bits should be stored in the event record 
available for the data analysis {\itshape (i)} to reduce the statistical
uncertainty of the efficiency determination (Sect.\,\ref{s:definitions}) and
{\itshape (ii)} to allow the weight calculation for fully efficient trigger
combinations (Sect.\,\ref{ss:advanced}).
\item
The optimum downscaling procedure should select events on a random basis,
to avoid end-of-run uncertainties (Sect.\,\ref{s:single}) and statistical
dependencies of (quasi-)identical trigger items (Sect.\,\ref{ss:incl1}).
\item
For deterministic downscaling systems,
several options to minimise the end-of-run correction exist:
{\itshape (i)} the status of the downscale counters at the end-of-run 
should be recorded; 
{\itshape (ii)} a randomly chosen position should be used for the 
selection in all downscale intervals of one run; 
{\itshape (iii)} the event in the middle of the downscale
interval should be selected (Sect.\,\ref{s:single}).
\item
The Inclusion Method assumes no correlation of the downscaling decisions
of different trigger items. For deterministic downscaling
systems, configurations with several \mbox{(quasi-)}identical trigger items
should be avoided. Alternatively the downscaling factors must fulfill certain
constraints (Sect.\,\ref{ss:incl1}).
\item
While downscale factors can be changed arbitrarily, frequent 
redefinitions of trigger items should be avoided. Every redefinition 
limits the run range in which the efficiency correction for the respective
trigger item must be determined (Sect.\,\ref{s:definitions})
and in which weight averaging for fully
efficient combinations of trigger items can be applied 
(Sect.\,\ref{ss:advanced}).
\item 
For an optimised trigger selection of events, sophisticated definitions
of trigger items combining many trigger elements might seem 
to be advantageous. However, very complex definitions should be avoided 
since the determination of their efficiency corrections
and correlations with other trigger items may be challenging 
(Sect.\,\ref{ss:eff}).
\item For multi-level trigger systems,
the simplest configuration for data analysis consists of parallel 1-to-1
chains (all-1-to-1-only).
If the assignment of several trigger items on one
level to the same trigger item on another level is unavoidable,
it should be restricted to separate 1-to-all, all-to-1 or all-to-all
configurations (Sect.\,\ref{s:highlevel}, especially \ref{ss:incl3}).
\item Although the final trigger decisions are based on the products of
actual trigger bits from different trigger levels,
the algorithms determining the raw trigger bits at higher levels
must not use the actual trigger bits from lower levels;
otherwise the Exclusion and Inclusion Methods for fully efficient trigger
combinations which involve raw trigger bits are inapplicable
(Sect.\,\ref{s:highlevel}).
\item
On all trigger levels the raw and actual bits of all trigger items 
used to select the analysed events should be available for the analysis
(see also Item\,1).
For early-accept systems this implies that 
the trigger information should be calculated in the offline data processing
where the selection code and the event parameters must be accessible
to reproduce all trigger decisions.
For early-reject systems the information
should be calculated either in the trigger system after a positive trigger
decision or likewise in the offline data processing (Sect.\,\ref{s:highlevel}).
\end{enumerate}

\section{Summary and Conclusions}\label{s:summary}

We have presented calculation methods for offline corrections of event
losses in trigger systems of particle collider experiments.
Emphasis has been put on the corrections of prescale factors and trigger
inefficiencies for combinations of event samples collected by different
trigger items.
Each method provides event weights, the sum of which reproduces the
original number of events that occured in the detector.
The methods have been discussed both for single-level and multi-level trigger
systems with and without considering uncorrelated and correlated
trigger inefficiencies.
We have studied the statistical performance of all methods and
considered implications for design and operation of trigger systems.

In detail, three conceptually different methods with increasing complexity
have been studied. The Division Method 
can provide sufficient statistical precision if the
individual trigger items have low downscale factors and high efficiencies
in their respective phase space regions. The accuracy can be
improved using the Exclusion Method which is adequate for many analyses.
The optimum performance is however provided by
the more complicated Inclusion Method which alone
makes use of all selected events in the combined sample.
For all methods the application of event weights averaged over run ranges
can yield a significant gain in the statistical precision of the result.

\section*{Acknowledgments}
\addcontentsline{toc}{section}{Acknowledgments}

This paper was inspired by work within the H1 collaboration.
Special acknowledgments belong to
the authors of the reports~\cite{h1-04-97-517,SchultzCoulon:1999tx}
who introduced the Inclusion Method for fully efficient trigger combinations,
as well as to V.\,Shekelyan who proposed the basic Exclusion Method
for one-level trigger systems.
We thank E.\,Elsen for useful comments and M.\,Medinnis
for proofreading this manuscript.
K.\,Kr\"uger is supported by the Bundesministerium f\"ur Bildung und
Forschung, Germany.


\begin{thebibliography}{00}
\addcontentsline{toc}{section}{References}

\bibitem{H1}
I.\,Abt {\itshape et al.} [H1 Collaboration],
\href{http://dx.doi.org/10.1016/S0168-9002(96)00893-5}%
{Nucl.\ Instrum.\ Meth.\ A {\bf 386} (1997) 310}. 

\bibitem{ZEUS}
ZEUS Collaboration,
Status Report,
Chapt. Trigger, DESY (1993),\\
URL: \href{http://www-zeus.desy.de/bluebook/bluebook.html}%
{http://www-zeus.desy.de/bluebook/bluebook.html}.

\bibitem{CDF}
CDF Collaboration,
Technical Design Report,
Chapt.\,12,
\ifthenelse{\boolean{twocolumn}}{\\}{}
\href{http://lss.fnal.gov/cgi-bin/find_paper.pl?pub-96-390-E}%
{Fermilab Pub-96/390-E} (1996).

\bibitem{D0}
D0 Collaboration,
Technical Design Report,
Part 3,
\ifthenelse{\boolean{twocolumn}}{\\}{}
\href{http://lss.fnal.gov/cgi-bin/find_paper.pl?pub-02-327-E}%
{Fermilab Pub-02/327-E} (2002).

\bibitem{ATLAS:2008zzm}
G.\,Aad {\itshape et al.}  [ATLAS Collaboration],
\href{http://dx.doi.org/10.1088/1748-0221/3/08/S08003}%
{JINST {\bfseries 3} (2008) S08003}, Chapt.\,8.

\bibitem{CMS:2008zzk}
R.\,Adolphi {\itshape et al.}  [CMS Collaboration],
\href{http://dx.doi.org/10.1088/1748-0221/3/08/S08004}%
{JINST {\bfseries 3} (2008) S08004}, Chapt.\,8.

\bibitem{Adloff:1997da}
C.\,Adloff {\itshape et al.}  [H1 Collaboration],
\href{http://dx.doi.org/10.1007/s100520050064}%
{Eur.\ Phys.\ J.\  C {\bfseries 1} (1998) 97}
\ifthenelse{\boolean{twocolumn}}{\\}{}
\mbox{[arXiv:\href{http://arXiv.org/abs/hep-ex/9709004}{hep-ex/9709004}].}

\bibitem{Adloff:2000qk}
C.\,Adloff {\itshape et al.}  [H1 Collaboration],
\href{http://dx.doi.org/10.1007/s100520100720}%
{Eur.\ Phys.\ J.\ C {\bfseries 21} (2001) 33}
\ifthenelse{\boolean{twocolumn}}{\\}{}
\mbox{[arXiv:\href{http://arXiv.org/abs/hep-ex/0012053}{hep-ex/0012053}].}

\bibitem{Adloff:2002ex}
C.\,Adloff {\itshape et al.}  [H1 Collaboration],
\href{http://dx.doi.org/10.1007/s10052-002-1009-8}%
{Eur.\ Phys.\ J.\  C {\bfseries 25} (2002) 25}
\ifthenelse{\boolean{twocolumn}}{\\}{}
\mbox{[arXiv:\href{http://arXiv.org/abs/hep-ex/0205064}{hep-ex/0205064}].}

\bibitem{excl}
V.\,Shekelyan, private communication.

\bibitem{h1-04-97-517}
S.\,Egli, E.\,Elsen, V.\,Lemaitre, K.\,M\"uller, H.\,Rick and
\ifthenelse{\boolean{twocolumn}}{\\}{}
\mbox{H.-C.\,Schultz-Coulon},
\ifthenelse{\boolean{twocolumn}}{}{\\}
{\mbox{H1 internal note 517}} (1997),
 unpublished.

\bibitem{SchultzCoulon:1999tx}
H.-C.\,Schultz-Coulon, J.\,Coughlan, E.\,Elsen, T.\,Nicholls and H.~Rick,\\
\href{http://dx.doi.org/10.1109/23.790703}%
{IEEE Trans.\ Nucl.\ Sci.\  {\bfseries 46} (1999) 915}.

\end{thebibliography}
\end{document}